\pdfoutput=1

\documentclass[10pt,aps,pre,twocolumn,superscriptaddress,footinbib]{revtex4-1}

\pdfoutput=1
\usepackage{graphicx}
\usepackage[left=1.5cm, right=1.5cm, top=1.785cm, bottom=2.0cm]{geometry}
\usepackage{amssymb}
\usepackage{moresize}
\usepackage{amsmath}
\usepackage{bm}
\usepackage{txfonts}
\hbadness=\maxdimen
\vbadness=\maxdimen
\usepackage{titlesec}
\usepackage[usenames]{color}

\def\be{\begin{equation}}
\def\ee{\end{equation}}
\def\bea{\begin{eqnarray}}
\def\eea{\end{eqnarray}}
\def\ba{\begin{array}}
\def\ea{\end{array}}

\begin{document}

\title{Normal stress anisotropy and marginal stability in athermal elastic networks}

\author{Jordan Shivers}
\affiliation{Department of Chemical and Biomolecular Engineering, Rice University, Houston, TX 77005, USA}
\affiliation{Center for Theoretical Biological Physics, Rice University, Houston, TX 77030, USA}
\author{Jingchen Feng}
\affiliation{Center for Theoretical Biological Physics, Rice University, Houston, TX 77030, USA}
\author{Abhinav Sharma}
\affiliation{Leibniz-Institut f\"{u}r Polymerforschung Dresden, 01069 Dresden, Germany}
\author{F.\ C.\ MacKintosh}
\affiliation{Department of Chemical and Biomolecular Engineering, Rice University, Houston, TX 77005, USA}
\affiliation{Center for Theoretical Biological Physics, Rice University, Houston, TX 77030, USA}
\affiliation{Departments of Chemistry and Physics \& Astronomy, Rice University, Houston, TX 77005, USA}

\begin{abstract}
Hydrogels of semiflexible biopolymers such as collagen have been shown to contract axially under shear strain, in contrast to the axial dilation observed for most elastic materials. Recent work has shown that this behavior can be understood in terms of the porous, two-component nature and consequent time-dependent compressibility of hydrogels. The apparent normal stress measured by a torsional rheometer reflects only the tensile contribution of the axial component $\sigma_{zz}$ on long (compressible) timescales, crossing over to the first normal stress difference, $N_1 = \sigma_{xx}-\sigma_{zz}$ at short (incompressible) times. While the behavior of $N_1$ is well understood for isotropic viscoelastic materials undergoing affine shear deformation, biopolymer networks are often anisotropic and deform nonaffinely. Here, we numerically study the normal stresses that arise under shear in subisostatic, athermal semiflexible polymer networks. We show that such systems exhibit strong deviations from affine behavior and that these anomalies are controlled by a rigidity transition as a function of strain. 
\end{abstract}


\maketitle

Normal solids and liquids exhibit shear stress under imposed shear deformation.  With the exception of simple Newtonian liquids, most materials also develop so-called normal stresses in response to shear.  Unlike shear stress, however, these stresses are directed perpendicular to surface on which they act and appear as diagonal terms in the stress tensor.  In the case of elastic solids, a common manifestation of normal stress is the Poynting effect, in which a solid tends to elongate in response to torsional strain.  In a classic series of experiments, Poynting observed such elongation for a variety of systems, ranging from simple metal wires to rubber \cite{Poynting1909,Poynting1913}.  By symmetry, this elongation should not depend on the sign or direction of the applied torsion, leading to lowest-order response that is expected to be quadratic in the the strain. This makes the Poynting effect a fundamentally nonlinear phenomenon.  This is one reason why normal stresses are typically less apparent than the shear stress, which varies linearly with strain.  Nevertheless, normal stresses have very dramatic consequences, including both rod climbing and tubeless siphoning, as well as die swell \cite{Larson1999}. These phenomena, as well as the Poynting effect, correspond to positive normal stress.

\begin{figure}[htp]
	\includegraphics[width=1.0\columnwidth]{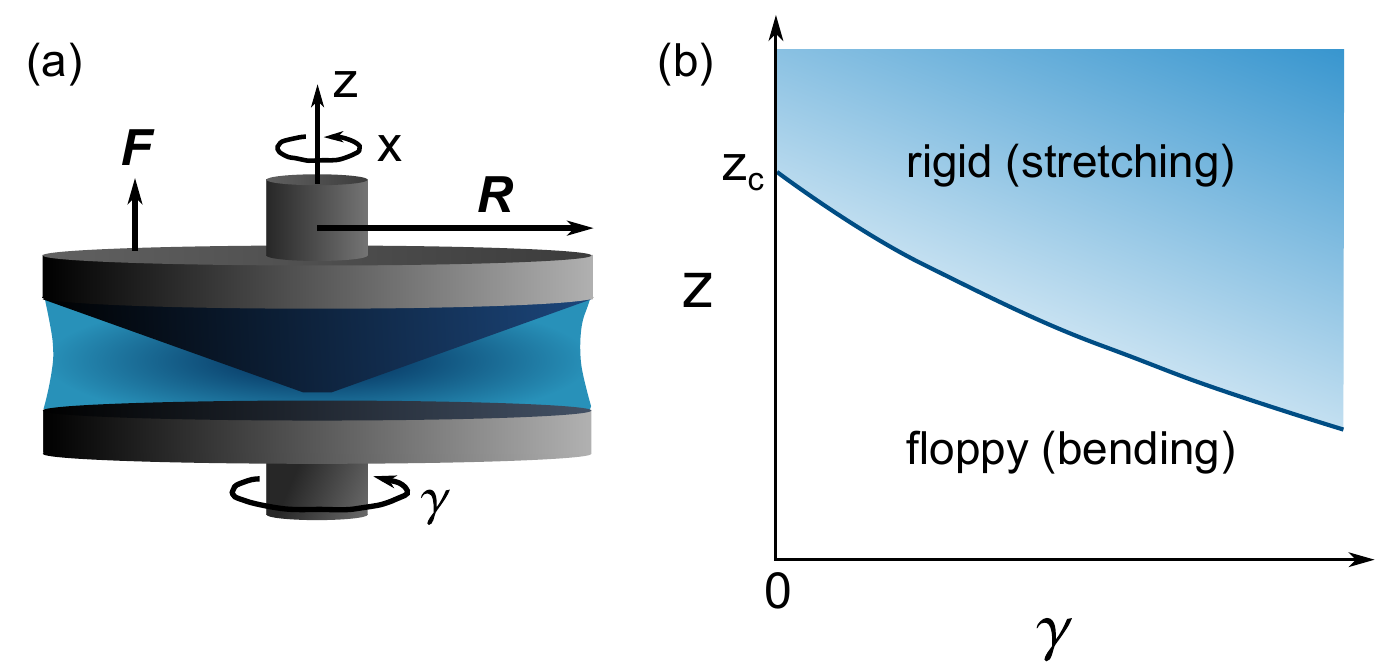}
	\caption{\label{fig1} (a) Schematic of a hydrogel sample in a cone-plate rheometer of radius $R$, with coordinates defined such that the $x$-axis and $z$-axis are oriented along the azimuthal (shear) and axial (gradient) directions, respectively. Positive axial force $F$ corresponds to the sample pushing up against the cone.   (b) Applying sufficient shear strain $\gamma$ to a subisostatic ($z<z_c$) network invokes a transition from a bending-dominated regime (floppy in the absence of bending interactions) below $\gamma_c$  to a stretching-dominated regime above $\gamma_c$. The details of the phase boundary $\gamma_c(z)$ (blue line) depend on the network structure. }
\end{figure}

In a cone-plate rheometer, shown schematically in Fig.\ \ref{fig1}a,  the measured axial force $F$ in torsion depends not only on the tensile axial stress component $\sigma_{zz}$, but also on the azimuthal component $\sigma_{xx}$, which acts as a hoop stress. For incompressible materials, this hoop stress generates a radial pressure gradient that contributes vertical thrust that counteracts $\sigma_{zz}$. In this case, the sign of the first normal stress difference, $N_1=\sigma_{xx}-\sigma_{zz}$ determines the sign of the measured axial force for sheared incompressible materials, according to $F = N_1\pi R^2/2$. The first normal stress difference, $N_1$, is fundamental to the nonlinear viscoelastic response of materials and is almost universally positive, particularly for solids.  For typical polymer networks, positive $N_1$ results from the fact that polymer extension in the azimuthal direction tends to be greater than in the axial direction \cite{Lodge1972}.  It was thus surprising when biopolymer gels, such as fibrin and collagen, were recently identified as apparent exceptions to this, with an inverted or negative Poynting effect \cite{Janmey2007, Kang2009}. Theory and simulation studies \cite{Janmey2007, Heussinger2007, Conti2009, Kang2009, Cioroianu2013, Licup2015, Meng2016, Licup2016} have shown that this observed negative normal stress is a generic feature of semiflexible networks, playing a significant role in the onset of the nonlinear strain-stiffening response characteristic of biopolymer networks \cite{Licup2015, Licup2016, Jansen2018}. However, as recently demonstrated \cite{DeCagny2016, Vahabi2017}, this anomaly for gels can be understood to arise from their porous, two-component nature. This porosity renders the gels effectively compressible on long enough time scales, over which the radial pressure gradient relaxes as the solvent flows from the sample boundaries, such that only the negative contribution from $\sigma_{zz}$ is measured, with $F = -\sigma_{zz}\pi R^2$ \cite{DeCagny2016, Vahabi2017}.  Consistent with this interpretation, these networks showed a normal (positive) Poynting effect on short enough time scales, in which the gels become effectively incompressible, indicating that the normal stress difference $N_1$ remains positive. 

For isotropic viscoelastic materials undergoing affine (homogeneous) simple shear deformation, the Lodge-Meissner relation relates $N_1$ to the shear stress $\sigma_{xz}$ as $N_1 = \sigma_{xz}\gamma$ \cite{Lodge1972}. This relation, first identified by Rivlin for elastic solids \cite{Rivlin1948}, holds for any material in which the principal strain axes and principal stress axes remain parallel throughout the applied deformation, which is satisfied as long as the material is initially isotropic and deforms affinely \cite{Larson1988}. Prior work has shown that networks of athermal fibers, of which collagen is a prime example, undergo highly nonaffine deformation under imposed shear strain. In such networks, it was recently shown that the degree of nonaffinity depends on the system's proximity to a strain-controlled transition that occurs along a critical line in the $\gamma$-$z$ plane, where $\gamma$ is the applied shear strain and $z$ is the connectivity, or average number of connections to each network junction \cite{Sharma2016a}. Strictly speaking, this transition occurs at finite strain for central-force networks below their isostatic point of (linear) marginal stability, as sketched in Fig.\ \ref{fig1}b. For fibers with finite bending rigidity, this line of marginal stability is manifest in a crossover from a soft, bending-dominated regime to a stiff, stretching-dominated regime. This nonlinear stiffening transition coincides with the development of a highly heterogeneous and anisotropic network of tensile force chains aligned primarily along the tension axis, similar to the marginally stable networks of compressive force chains that develop at the jamming transition in sheared granular packings \cite{Majmudar2005, Voivret2009} and frictional force chains in shear-thickening suspensions \cite{Melrose2004, Lin2016}, both of which align instead along the compression axis. While force chains have been observed in fibrous networks \cite{Heussinger2007a, Amuasi2015, Liang2016, Ronceray2016}, the properties of force chain networks that develop during macroscopic strain stiffening, and their effects on the normal stresses, have not been extensively studied. In shear-thickening suspensions, the formation of such force networks are typically associated with anomalous, and sometimes negative, values of $N_1$ \cite{Mari2014,Cwalina2014, Seto2018}. Given the similarity of the force chains in sheared semiflexible fiber networks to those observed in packings/suspensions, as well as the significant nonaffinity observed near the strain-stiffening transition, it is not obvious that the Lodge-Meissner relation should apply, or even that $N_1$ should be positive for these networks. While some prior theory and simulation \cite{Cioroianu2013, Amuasi2015} studies have suggested that semiflexible networks may generally satisfy the Lodge-Meissner relation, a systematic study of the effects of network structure and nonaffinity on $N_1$ has been lacking. 

Here, we investigate the behavior of the various normal stress components in athermal subisostatic fiber networks near the strain-stiffening transition, using numerical models of disordered semiflexible fiber networks in two and three dimensions. We show that the general scaling of the normal stresses with shear strain below, near, and above the rigidity transition remains consistent irrespective of the underlying network structure. However, we demonstrate that such networks can exhibit anomalous behavior in $N_1$ that is highly sensitive to the network structure, and that this anomaly is most pronounced near the point of marginal stability as a function of strain, i.e., along the phase boundary in Fig.\ \ref{fig1}b that corresponds to nonlinear strain-stiffening. This anomaly at the stiffening transition results from the formation of a highly heterogeneous, anisotropic, system-spanning network of strong tensile force chains, whose spatial structure and force distribution determines the relative values of each normal stress component and thus $N_1$. Our results suggest that any underlying anisotropy in the network structure can result in anomalous behavior in $N_1$ that is maximized at the critical strain, suggesting that the sign and magnitude of $N_1$ can, in principle, be tuned by selectively modifying the network structure.  Interestingly, our results suggest that in the limit of very large and nearly isotropic systems, such as large off-lattice network models or experimental gels, the Lodge-Meissner relation should be satisfied at any strain, in spite of the significant nonaffine deformations and heterogeneous force network associated with the critical strain. 

\begin{figure}[htp]
\includegraphics[width=1.0\columnwidth]{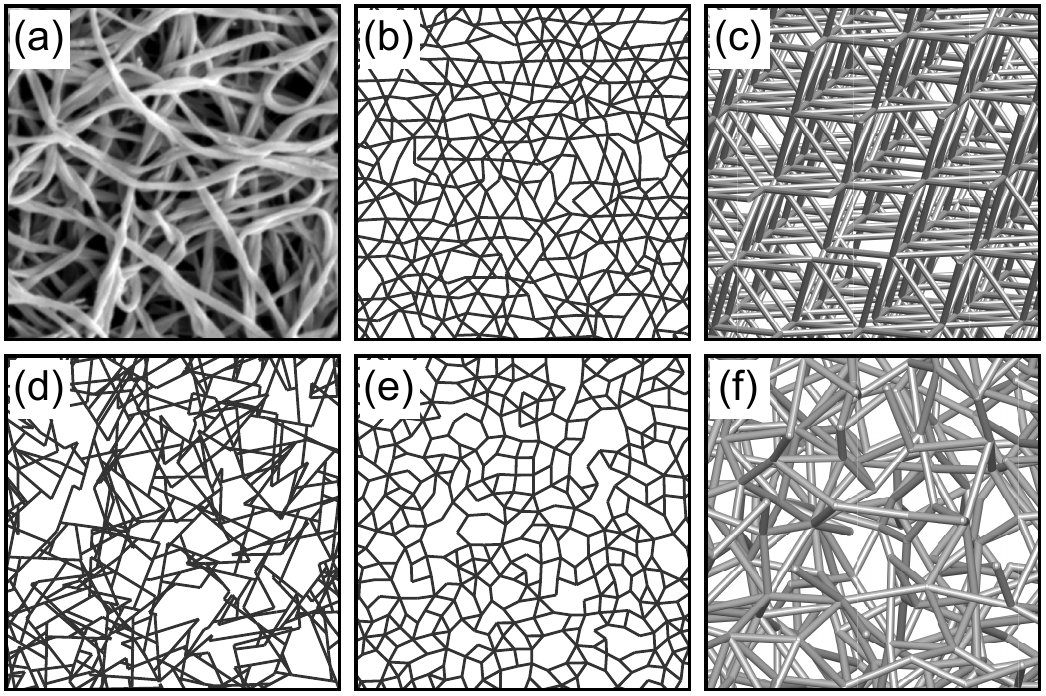}
\caption{\label{fig2}
	(A) Sample of a reconstituted collagen network exhibiting clear connective and geometric disorder, adapted from Ref.\ \cite{Sharma2016a}. We investigate the mechanics of bond-diluted athermal semiflexible fiber networks including (b) phantom 2D triangular networks with added positional disorder, (c) phantom FCC lattice-based networks, (d) random fiber (Mikado) networks, (e) 2D bidisperse disk packing-derived networks and (f) 3D bidisperse sphere packing-derived networks.}
\end{figure}

\section*{Numerical models}

We consider discrete models of semiflexible polymer networks in 2 and 3 dimensions, including both lattice-based and off-lattice network structures, with filament-bending (i.e. freely hinging crosslinks between fibers) and bond-bending interactions. For lattice-based models, we consider two-dimensional (triangular) and three-dimensional (face-centered cubic) lattice-based networks, and for off-lattice networks we consider two-dimensional Mikado and bidisperse disk packing-derived networks as well as three-dimensional bidisperse sphere packing-derived networks. Examples of these are shown in Figure \ref{fig2}. 

We construct disordered lattice-based networks in 2D beginning with fibers arranged on a periodic triangular lattice with lattice spacing $l_0 = 1$ and sides of $W$ lattice units \cite{Thorpe1985,Das2007}, which we then \emph{phantomize} by disconnecting one of three intersecting fibers at each node, in order to reduce the average network connectivity $\langle z \rangle$ to 4 \cite{Broedersz2011a}. Prior work has shown that, in 2D networks, the buckling of long, straight fibers leads to unrealistic mechanical effects including a dip in the differential shear modulus $K=\partial\sigma_{xz}/\partial\gamma$ \cite{Licup2016}. We avoid this by introducing geometric distortion to the unstrained lattice network by moving each node a random distance in the range $[0,\delta_{\rm max}]$ in a random direction, with $\delta_{\rm max}\le0.5$ in order to avoid overlapping nodes \cite{Jacobs1996,Onck2005,Rens2016}, and subsequently redefining the rest lengths $l_{ ij,0}$  between pairs of nodes and rest angles $\theta_{ijk,0}$ between connected triplets of nodes so that the geometrically disordered network exhibits zero stress in the unstrained state. In order to avoid system-spanning (or nearly system-spanning) fibers, which introduce unrealistic contributions to the macroscopic mechanics \cite{Broedersz2011, Licup2015}, we remove every $q$th bond along each fiber, beginning with a randomly chosen bond, prior to dilution. For phantomized triangular networks, we use $W = 120$ and $q = 20$. Similarly, the process for generating 3D face-centered cubic (FCC) lattice-based networks (as shown in Figure $\ref{fig2}$c) begins with fibers arranged on a periodic FCC lattice \cite{Feng1984} with sides of $W$ lattice units with lattice spacing $l_0 = 1$. We phantomize these as well \cite{Broedersz2012}, yielding an average $z = 4$, and cut a single randomly chosen bond on each fiber prior to dilution.  We use 3D FCC networks with $W = 25$.

Mikado networks are constructed by placing straight segments of length L with random positions and orientations into a 2D periodic box of side length $W$, adding crosslinks at the intersections between segments \cite{Wilhelm2003, Head2003}. Fibers are deposited randomly until the desired average crosslink density $L/l_c$ is reached, where $l_c$ is the average bond length. Even in the infinite crosslink density limit, Mikado networks yield $z < 4$. We generate networks with $W = 10L$ and $L/l_c\approx12$, yielding an initial connectivity of $z \approx 3.6$ prior to dilution. 

We prepare 2D packing-derived (PD) networks by first randomly placing $N = W^2$ radially bidisperse disks with harmonic repulsive interactions within a periodic square unit cell of side length $W$, where half of the disks are assigned a radius $r = r_0$ and half $r = \phi r_0$, with $\phi = 1.4$ chosen to avoid long-range order \cite{Koeze2016}. We incrementally increase $r_0$ from $0$ until the system jams, exhibiting a finite bulk modulus. From this disordered packing, we generate a contact network by connecting the centers of the overlapping disks (excluding rattlers) with springs at their rest lengths \cite{VanHecke2010, Ohern2003, Dagois-Bohy2012}. The same procedure is followed in 3D, using $N = W^3$ radially bidisperse harmonic repulsive spheres, also with $\phi = 1.4$, in a periodic cubic unit cell of side length $W$. With sufficiently large systems, this procedure generates contact networks with $z \approx 2d$, where $d$ is the dimensionality.  Unless otherwise stated, we study 2D packing-derived networks with $W = 100$ ($N = 10000$ nodes) and 3D packing-derived networks with $W = 20$ ($N = 8000$ nodes). 

After generating the underlying network structure, we repeatedly remove randomly chosen bonds and any consequent dangling ends until the network reaches the desired average network connectivity $z$. We model the lattice-based and Mikado networks as filamentous networks with freely-hinging crosslinks, in which bending interactions are accounted for only along each fiber \cite{Head2003, Sharma2016a}, whereas the packing-derived networks are modeled instead as bond-bending networks \cite{Arbabi1988} with bending interactions between all pairs of nearest-neighbor bonds. Given that the precise mechanics of the connections between fibers in real collagen and fibrin networks, which can include both branching points and crosslinks, are not well characterized, using two different crosslink models enables us to study whether the behavior of the normal stresses is independent of the detailed form of the bending interactions. Prior work has shown that these models exhibit similar linear mechanics \cite{Broedersz2011, Das2012} and strain-driven critical behavior \cite{Sharma2016a, Sharma2016, Rens2016}. Energetically, we treat individual bonds as Hookean springs with stretching modulus $\mu$ and pairs of bending-associated bonds with bending modulus $\kappa$. 
The Hamiltonian $\mathcal{H}$ of the full network is
\begin{equation}
\mathcal{H} = \frac{1}{V} \left[\dfrac{\mu}{2}\sum_{\langle ij\rangle}{\dfrac{\left(l_{ ij}-l_{ ij,0}\right)^2}{l_{ij,0}}} + \dfrac{\kappa}{2}\sum_{\langle ijk\rangle}{\dfrac{\left(\theta_{ ijk}-\theta_{ ijk,0}\right)^2}{l_{ ijk,0}}}\right],
\end{equation}
in which the sums are taken over pairs $\langle ij \rangle$ and triplets $\langle  ijk \rangle$ of connected nodes, and $l_{ijk,0} = (l_{ij} + l_{jk})/2$. For networks with freely hinging crosslinks, the second sum is taken only for adjacent pairs of bonds along fibers. Here, $V = v_0W^d$, where $v_0 = \sqrt{3}/2$ for triangular lattice-based networks, $v_0 = \sqrt{2}/2$ for FCC lattice-based networks, $v_0 = 1$ otherwise, and $d$ is the dimensionality. As in prior work, we set $\mu=1$ and define a dimensionless bending rigidity $\tilde{\kappa} = \kappa/\mu l_c^2$. Bond-diluted network models such as these have been shown to quite effectively describe the shear elasticity of reconstituted collagen networks \cite{Sharma2016a, Jansen2018}, which have a typical average value of $ z  \approx 3.4$ \cite{Lindstrom2010}. 

We perform simulations of networks under simple shear by incrementally increasing the shear strain $\gamma$ from  $10^{-2}$ to $1$ in exponentially spaced steps, using generalized Lees-Edwards periodic boundary conditions \cite{Lees1972}. For simplicity of notation when comparing 2D and 3D simulations, we denote $x$ and $z$ the directions of shear and gradient, respectively, in both cases.  At each strain value, the network energy is minimized using the FIRE algorithm \cite{Bitzek}, and each component of the stress tensor $\bm{\sigma}$ is computed as
\begin{equation}
\sigma_{\alpha\beta} = \dfrac{1}{2V}\sum_{\langle ij\rangle}{f_{ij,\alpha}u_{ij,\beta}}
\end{equation}
in which $\mathbf{u}_{ij} = \mathbf{u}_j -\mathbf{u}_i$ is the vector between nodes $i$ and $j$ and $\mathbf{f}_{ij}$ is the force acting on node $i$ due to node $j$ \cite{DoiEdwards}. To symmetrize the normal stresses in the linear regime, we average the response of each network sample under positive and negative shear strain. Unless otherwise stated, the reported stress is averaged over at least 10 samples. 

For comparison, we also consider the limit of an isotropic medium composed of filaments of length $l_0 = 1$ with uniformly distributed initial orientations, which are assumed to deform affinely under simple shear. We compute the resulting stress tensor as a function of strain for this system with two complementary force extension relations: simple linear Hookean springs, which support both compression and tension, and ``rope''-like springs that support only tension (see Supplementary Information). For the remainder of the paper we refer to the first isotropic model as the spring model and the second as the rope model. Both analytical models satisfy the Lodge-Meissner relation under any applied strain.

\section*{Normal stresses and strain-stiffening}

\begin{figure}[ht]
\includegraphics[width=1.0\columnwidth]{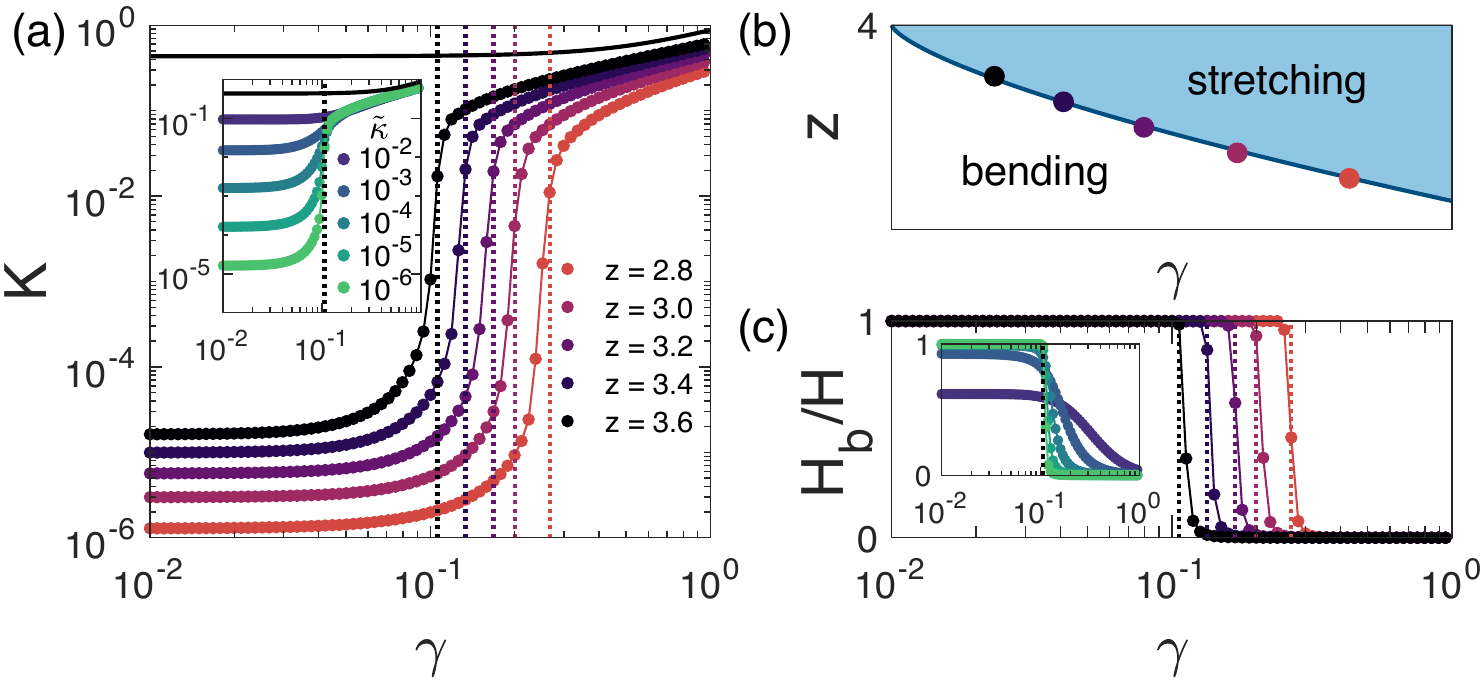}
\caption{\label{fig3}  (a) Differential shear modulus $K$ vs. strain $\gamma$ for phantomized triangular networks with $\tilde{\kappa} = 10^{-6}$, $W = 120$, $\delta_{\rm max}=0.4$, and varying connectivity $z$. The vertical dotted line for each $z$ value indicates the critical strain $\gamma_c$, determined as the strain corresponding to the onset of finite $K$ in the low-$\tilde{\kappa}$ limit. These curves illustrate that the critical strain increases with decreasing $z$.  The solid black line shows the computed $K$  for the affine isotropic network model with line density $\rho = 2\sqrt{3}$.  Inset: $K$ vs. $\gamma$ for constant $z = 3.6$ and varying $\tilde{\kappa}$.  (b) Schematic mechanical phase diagram indicating the increase in $\gamma_c$ with decreasing $z$ below $z_c = 2d$. (c) For the same networks, the ratio of bending energy $\mathcal{H}_b$ to total energy $\mathcal{H}$ illustrates the bending-to-stretching transition that occurs at the critical strain and (inset) sharpens with decreasing $\tilde{\kappa}$. Colors in both the main panel and inset correspond to those in panel (a). Lines between points are intended to serve as guides to the eye.}
\end{figure}

Without bending interactions, spring networks exhibit a finite linear shear modulus $G = \lim_{\gamma\to0} K>0$ only when their connectivity $z$, defined as the average number of connections at each node, reaches a critical isostatic connectivity $z_c$ \cite{Broedersz2011}. While the precise value of $z_c$ is sensitive to the heterogeneity of the network structure, typical values are close to the constraint-counting value $z_{iso}=2d$ introduced by Maxwell \cite{Maxwell1864}. Under shear strain, spring networks that are subisostatic, with $z < z_c$, develop finite $K\approx\mu$ at a critical strain $\gamma_c$ that depends on the network's connectivity and geometry, with $\gamma_c \to 0$ as $z\to z_c$ from below. At the critical strain, such networks develop a system-spanning branched network of primarily tensile force chains, oriented predominantly along the principal extension axis, in order to support finite stress. Associated with the development of this force chain network are characteristic signatures of criticality including diverging nonaffine fluctuations \cite{Sharma2016a}.  In networks with finite $\tilde{\kappa}$, $K$ is finite and proportional to $\tilde{\kappa}$ below the critical strain, and subisostatic semiflexible polymer networks therefore undergo a transition from a bending-dominated regime to a stretching-dominated regime at the critical strain \cite{Sharma2016a}.  In Fig.\ \ref{fig3}, we show $K$ vs. $\gamma$ for several values of $z$, demonstrating that $\gamma_c$ increases with decreasing $z$. For constant $z$, $\gamma_c$ is very weakly dependent on $\tilde{\kappa}$ in the limit of $\tilde{\kappa}\to0$, and the networks exhibit a clear transition from a bending dominated regime ($K\propto\tilde{\kappa}$)  for $\gamma<\gamma_c$  to a stretching dominated regime ($K\propto\mu$) for $\gamma>\gamma_c$ (Fig.\ \ref{fig3}a inset). This behavior is also clear from the proportion of the total energy arising from bending interactions, $\mathcal{H}_b/\mathcal{H}$, as we show in Fig.\ \ref{fig3}c: as $\tilde{\kappa}$ is decreased, the transition from the bending-dominated to stretching-dominated regime at $\gamma_c$ sharpens. One can map the critical strain as a function of $z$ to yield a phase diagram for the mechanical behavior of subisostatic networks as a function of strain and connectivity, as shown schematically in Fig \ref{fig3}b \cite{Sharma2016a}.  The details of the phase boundary depends on the underlying network geometry. 

\begin{figure}[ht]
\includegraphics[width=1.0\columnwidth]{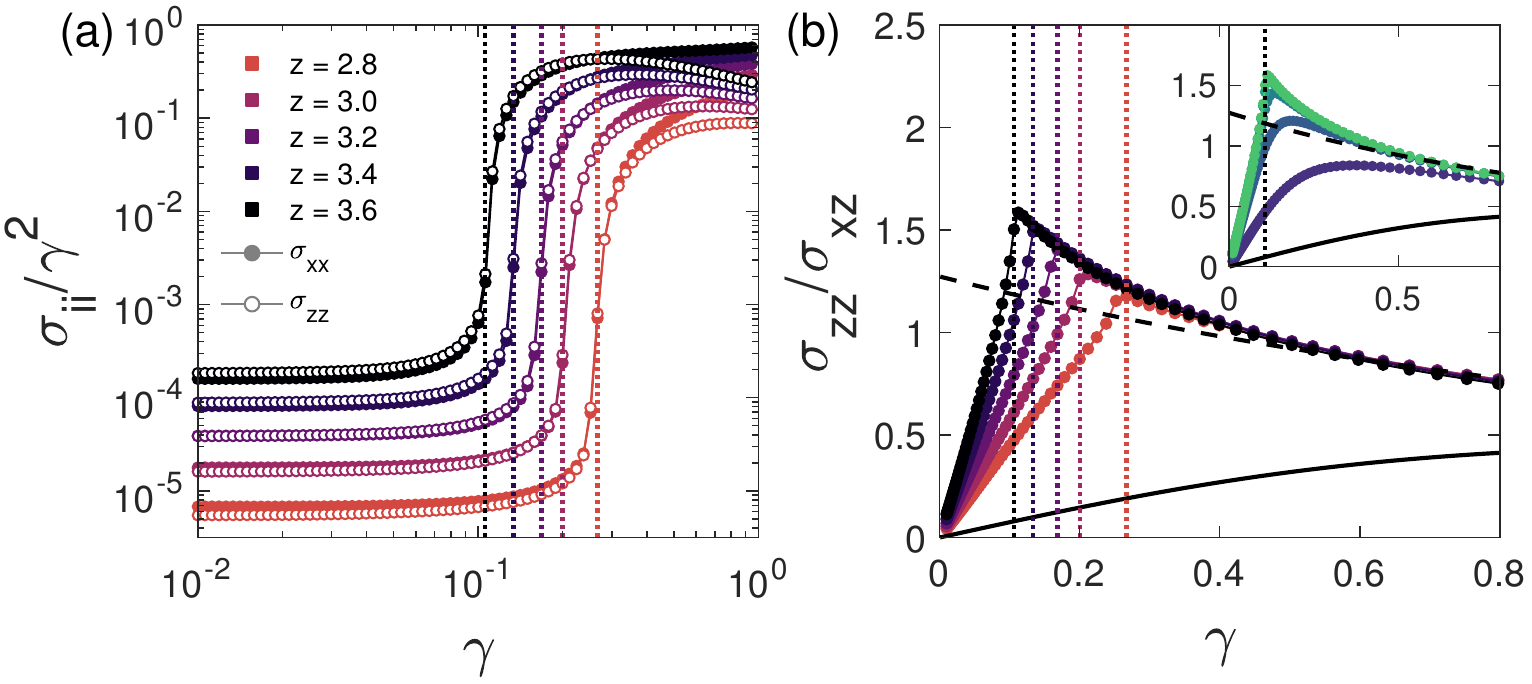}
\caption{ \label{fig4} Normal stress components $\sigma_{xx}$ (closed symbols) and $\sigma_{zz}$ (open symbols) normalized by $\gamma^2$, for the same phantom triangular networks as in Fig.\ \ref{fig3}, with $\tilde{\kappa} = 10^{-6}$ and varying $z$. Dotted lines indicate $\gamma_c(z)$. (b) The corresponding ratio of the axial normal stress $\sigma_{zz}$ to the shear stress $\sigma_{xz}$ shows a peak at the critical strain for each $z$ value that (inset) sharpens with decreasing $\tilde{\kappa}$. Colors correspond to those in Fig.\ \ref{fig3}. The thick solid line corresponds to the affine isotropic spring network model, and the dashed solid line corresponds to the affine isotropic rope network model.}
\end{figure}

\begin{figure}[ht]
\includegraphics[width=1.0\columnwidth]{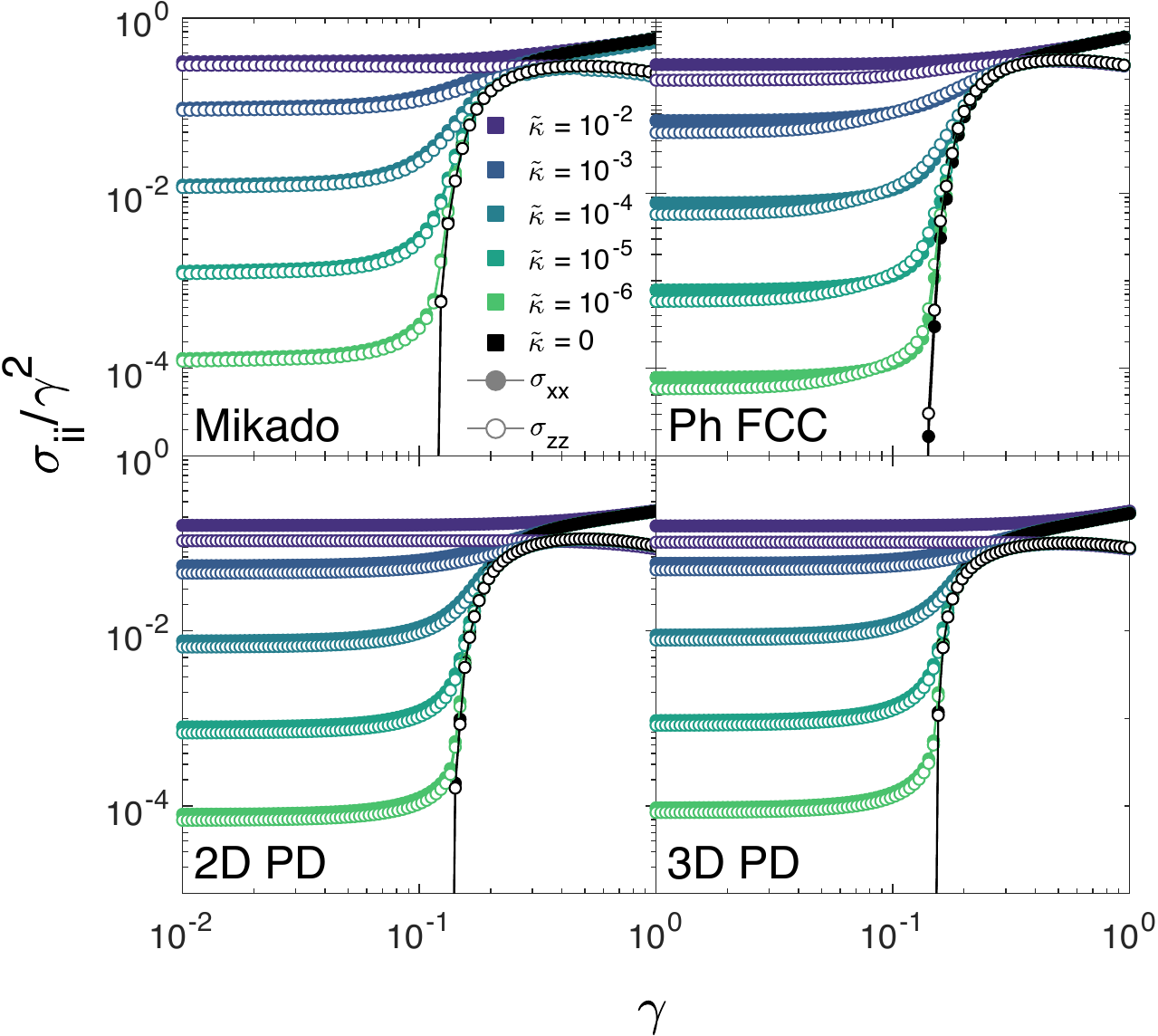}
\caption{ \label{fig5} Normal stress components $\sigma_{xx}$ (closed symbols) and $\sigma_{zz}$ (open symbols) normalized by $\gamma^2$ for (top left) Mikado networks with $z = 3.3$,  (top right) phantom FCC networks with $z = 3.4$, (bottom left) 2D packing-derived networks with $z = 3.3$, and (bottom right) 3D packing-derived networks with $z = 4.8$,  all with varying $\tilde{\kappa}$. }
\end{figure}

The normal stress components $\sigma_{ii}$ (where $i = x, z$) both exhibit the same bending-dominated to stretching-dominated transition at the $z$-dependent critical strain, with $\sigma_{ii}\propto \tilde{\kappa}$ for $\gamma<\gamma_c$ and $\sigma_{ii}\propto\mu$ for $\gamma > \gamma_c$, as shown in Fig.\ \ref{fig4}a for phantom triangular networks. As we show in Fig.\ \ref{fig5}, we observe essentially the same behavior in all subisostatic network models studied here, supporting the idea that the details of the network structure \cite{Sharma2016a} and bending energy type \cite{Rens2016} have only minor effects on the general strain-stiffening behavior of semiflexible fiber networks. Instead, the governing variables are $z$ and $\gamma$. In Supplementary Information, we show that phantomized triangular networks exhibit the same mechanical behavior with freely-hinging crosslinks as with bond-bending interactions, with the only difference being that bond-bending interactions leads to a higher apparent $\kappa$ due to the additional angle constrains.  We also observe that the ratio of the axial normal stress component $\sigma_{zz}$ to the shear stress $\sigma_{xz}$ becomes maximal, and typically greater than 1, at $\gamma_c$, as we show in Fig.\ \ref{fig4}b. This peak reflects the sharp transition from the bending-dominated regime, in which $\sigma_{zz}\propto\gamma^2$ and $\sigma_{xz}\propto\gamma$ for small $\gamma$, yielding $\sigma_{zz}/\sigma_{xz}\propto\gamma$, to the stretching-dominated regime, in which $\sigma_{xz}$ grows as a power-law with respect to $\Delta\gamma$ and rapidly begins to dominate $\sigma_{zz}$. Preliminary observations of this phenomenon were made in prior work using Mikado networks \cite{Conti2009} and in experiments on fibrin \cite{Kang2009}. We observe, in all network models discussed here, that the peak follows the critical strain as $z$ is varied and grows to a $z$-dependent asymptotic value as $\tilde{\kappa}$ decreases. A maximum in this ratio actually does occur in the affine isotropic spring network limit due to the gradual reorientation of fibers under increasing shear strain, but the maximal value is smaller ($\sim0.4$) and the peak strain much larger ($\gamma\sim1$) than we observe for our semiflexible network models. The isotropic rope network model, in contrast, exhibits a maximum with $\sigma_{zz}/\sigma_{xz}>1$ at zero strain. In semiflexible networks with small $\tilde{\kappa}$, it appears that the critical strain marks a transition from the linear, bending-dominated regime in which $\sigma_{zz}/\sigma_{xz}\propto\gamma$ to a rope-like regime. This results from the fact that, at and above the critical strain, tension forces vastly outweigh compressional forces in networks with low $\tilde{\kappa}$.  In Fig.\ \ref{fig4}b, we show that systems with low $\tilde{\kappa}$ exhibit a ratio $\sigma_{zz}/\sigma_{xz}$ which, above $\gamma_c$, is quite close to the rope network limit, whereas for higher $\tilde{\kappa}$ networks the ratio approaches the spring network limit. The latter trend is expected, as increasing $\tilde{\kappa}$ increases the degree to which the network's fibers can support compression. Interestingly, we observe that, near the critical strain, $\sigma_{zz}/\sigma_{xz}$ exceeds even the rope values predicted for the affine isotropic rope network limit, possibly due to the highly heterogeneous nature of the stress-bearing network at the critical strain. Nevertheless, it is apparent that a large ratio of the axial normal stress to the shear stress is a signature of the development of a rope-like stress-bearing structure at the critical strain. This is further supported by prior experimental evidence that fibrin networks with stiffer filaments exhibit a smaller peak in $\sigma_{zz}/\sigma_{xz}$ than more flexible ones at the critical strain \cite{Kang2009}.

\section*{Stress anisotropy}

For typical isotropic elastic materials, the first normal stress difference $N_1=\sigma_{xx}-\sigma_{zz}$ is positive, and for affinely deforming isotropic elastic materials like rubber it is typically well approximated by the Lodge-Meissner relation $N_1 = \sigma_{xz}\gamma$. Negative values of $N_1$ are unusual, but have been observed in certain materials including shear-thickening suspensions \cite{Cwalina2014}. However, negative $N_1$ has not to date been observed in a real elastic solid. While the normal stresses we observe for all networks (Figs. \ref{fig4} and \ref{fig5}) are similar in magnitude at and below the critical strain, we observe that the behavior of the first normal stress difference $N_1$ depends strongly on the underlying network structure. 

\begin{figure}[ht]
\includegraphics[width=1.0\columnwidth]{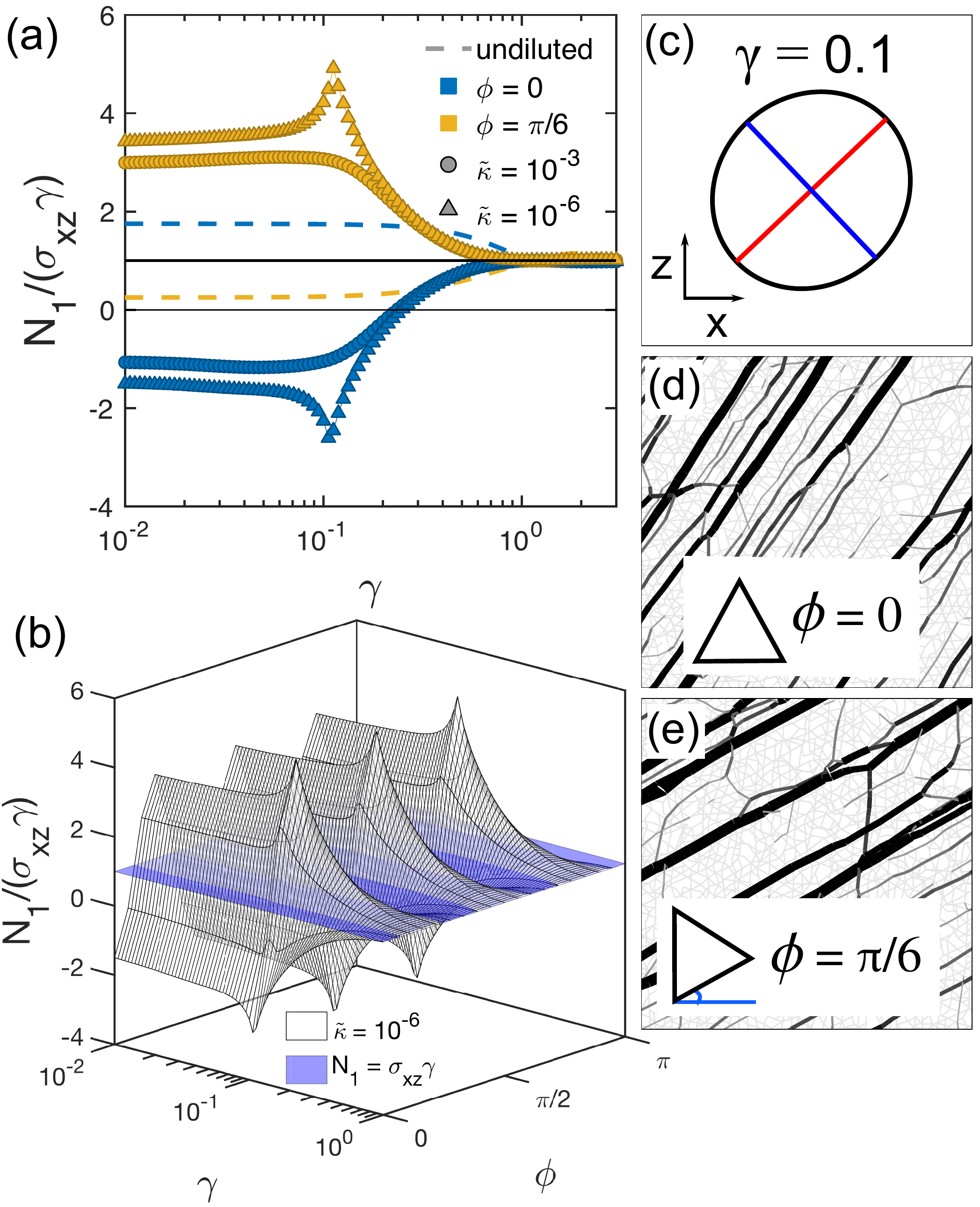}
\caption{\label{fig6} (a) $N_1$ normalized by the Lodge-Meissner relation $N_1 = \sigma_{xz}\gamma$ for undiluted triangular lattices (dashed lines) and phantomized triangular networks with $z = 3.6$, and $W = 120$, and varying $\tilde{\kappa}$, in which the lattice is initially rotated by angle $\phi = 0$ and $\phi = \pi/6$. In the subisostatic lattice case, the peak at the critical strain $\gamma_c$ changes sign when the lattice is rotated by $\phi = n\pi/6$ with odd $n$, as shown in (b) for $\tilde{\kappa} = 10^{-6}$ over the full range of $\theta$. In lattice-based networks such as these with long fibers along specific lattice vectors, force chains preferentially develop at the critical strain along whichever lattice vector is closest to the principal extension direction for a given applied strain $\gamma$ (see Supplementary Information). (c) The principal extension axis for the simple shear deformation gradient $\Lambda(\gamma=0.1)$ is shown in red, with the (perpendicular) principal compression axis shown in blue. In black, we show the corresponding strain ellipsoid, projected onto the $x$-$z$ plane. d) The angular orientation of the dominant force chains relative to $\theta=\pi/4$ determines the sign of $N_1$. For unrotated networks ($\phi= 0$) with the same parameters as in (b), the most elongated fibers at the critical strain $\gamma_c\approx0.1$ are oriented with $\theta_b>\pi/4$, resulting in negative $N_1$. (e) When these networks are rotated initially by $\phi = \pi/6$, the most elongated fibers at the critical strain are instead oriented with $\theta_b<\pi/4$, resulting in positive $N_1$.}
\end{figure}

Under applied strain $\gamma$, the periodic images of each node in the network transform affinely according to the simple shear deformation gradient $\Lambda(\gamma)$. This deformation gradient results in maximal elongation along its principal extension axis with orientation $\theta_P$, with maximal compression along the perpendicular axis, as shown schematically in Fig.\ \ref{fig6}c for a small strain. We determine the principal strains and principal strain axes, which rotate with applied strain, as a function of $\gamma$ in Supplementary Information.  For $\gamma = 0$, the principal extension axis is oriented $\theta_P=\pi/4$ radians above the $x$-axis in the $x$-$z$ plane. For isotropic and affinely deforming networks with only axial forces, such as the rope and spring model, the principal stress axes exactly follow the principal strain axes under any applied $\gamma$, such that the Lodge-Meissner relation is always satisfied. That the principal stress axes follow the principal strain axes is not guaranteed in disordered networks, as they deform nonaffinely and are not perfectly isotropic. 

In particular, lattice-based networks such as triangular and FCC models exhibit significant angular anisotropy; in the unstrained state, their bonds lie only along vectors corresponding to the lattice directions, and imposed local geometric disorder does little to mitigate this long-range anisotropy. We can explore the effects of this anisotropy by applying an initial rotation of $\phi$ radians, relative to the $x$-axis in the $x$-$z$ plane, to the lattice prior to applying shear strain. Arbitrarily, we define the unrotated ($\phi=0$) phantom triangular lattice as having bonds initially oriented at angles $\theta_{b,0} \in [0, \pi/3, 2\pi/3]$ relative to the $x$-axis. Even with significant random local geometric distortion $\delta_{max} = 0.4$, the fibers remain \textit{on average} oriented along these initial lattice vectors. In general, tensile force chains develop in randomly diluted spring networks at the critical strain and tend to be oriented along the principal extension axis. As bonds in a phantom triangular network do not have a uniform initial angular distribution and are instead oriented primarily along the initial lattice bond orientations for a given $\phi$, the tensile force chains develop along the (slightly rotated) initial lattice bond orientation that is most stretched at $\gamma_c$, i.e. whichever is closest to the principal extension axis.

For initially unrotated ($\phi = 0$) triangular lattice networks, the dominant tensile force chains develop primarily along the bonds that are initially oriented along the $\theta_{b,0} = \pi/3$ direction, as shown in Fig.\ \ref{fig6}d, with softer branches oriented along the other directions. As a result, the maximum principal stress is oriented close to the $\theta_{b,0} = \pi/3$ direction, not parallel to the principal extension axis. The initial lattice orientation determines which of the (rotated) initial lattice bond orientation the tensile force chains propagate along at $\gamma_c$, thus determining the relative orientation of the maximal principal stress axis to the principal extension axis. When the initial lattice is rotated by $\phi = \pi/6$, such that the initial undistorted lattice bond vectors are oriented along  $\theta_{b,0} \in [\pi/6, \pi/2, 5\pi/6]$, the dominant force chains instead propagate along the rotated lattice vector corresponding to  $\theta_{b,0}=\pi/6$, as shown in Fig.\ \ref{fig6}e.

But how does this affect $N_1$? In the $\tilde{\kappa} = 0$ limit, in which forces only occur parallel to bonds, the value of $N_1$ is entirely determined by the individual bond orientations, with $N_1 \propto \sum_b f_b l_b \cos(2\theta_b)$ where $f_b$ is the tension, $l_b$ is the length, and $\theta_b\in[-\pi/4,3\pi/4]$ is the angle of bond $b$ relative to the $x$-axis in the $x$-$z$ plane. This range for $\theta_b$ is convenient, as bonds under tension with $\theta_b > \pi/4$ exhibit negative $N_1$, whereas bonds under tension with $\theta_b < \pi/4$ exhibit positive $N_1$. A similar expression was used in Ref. \cite{Seto2018} to describe relative contributions to $N_1$ based on force networks in non-Brownian suspensions. With finite $\tilde{\kappa}$, forces also occur perpendicular to bonds, leading to a more complicated dependence of $N_1$ on the network configuration. Since the tensile force networks dominate for relatively low-$\tilde{\kappa}$ networks at and above the critical strain, it is reasonable to estimate $N_1$ for such networks in this regime only in terms of stretching forces, i.e. as a simple function of the bond orientations. 

\begin{figure}[ht]
\includegraphics[width=1.0\columnwidth]{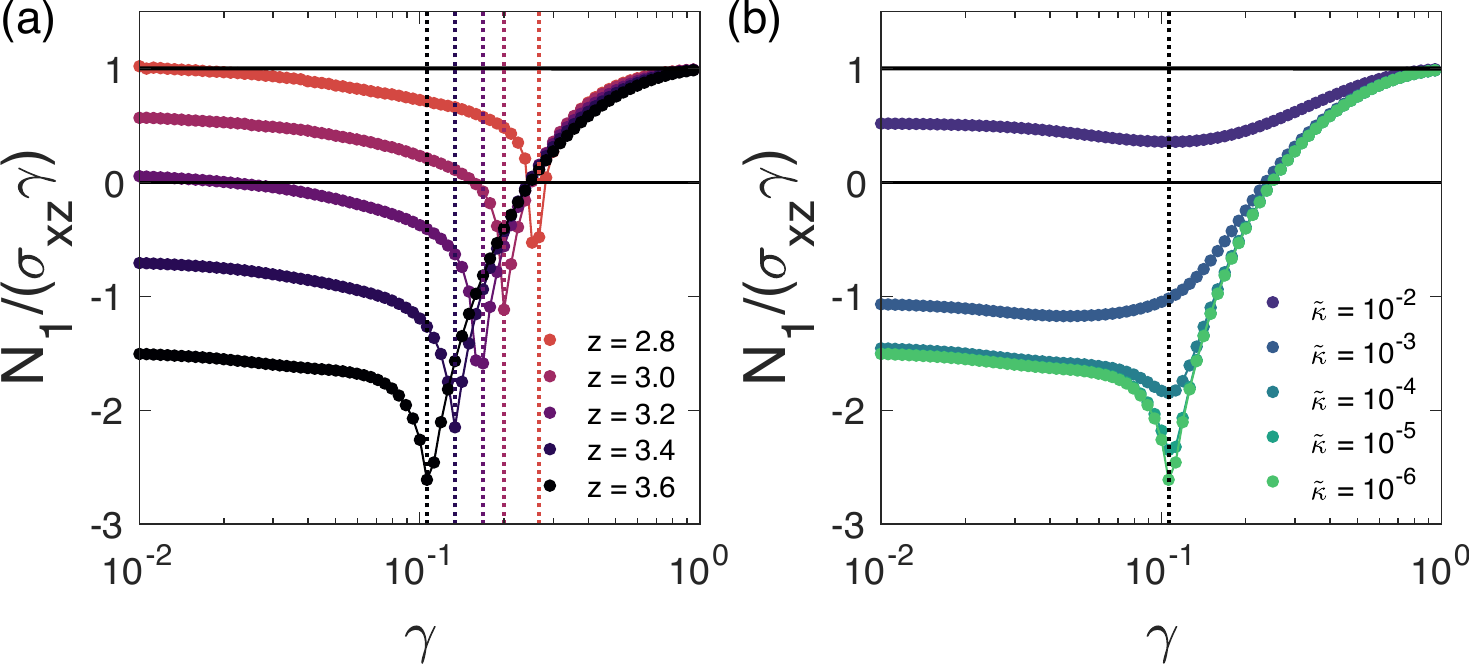}
\caption{ \label{fig7} First normal stress difference $N_1\equiv\sigma_{xx}-\sigma_{zz}$ normalized by the Lodge-Meissner relation ($N_1 = \sigma_{xz}\gamma$) in unrotated 2D distorted phantomized triangular networks ($\phi = 0$, $\delta_{\rm max} = 0.4$) as a function of $\gamma$ for (a) $\tilde{\kappa} = 10^{-6}$ and varied $\langle z \rangle$ and (b) $\langle z \rangle = 3.6$ and varied $\tilde{\kappa}$, with the Lodge-Meissner result ($N_1/(\sigma_{xz}\gamma)=1$) shown as a thick solid line. Deviation from the Lodge-Meissner relation increases with $\langle z \rangle$, and the ratio exhibits a downward peak and maximal anomaly at the critical strain $\gamma_c$, which grows with decreasing $\tilde{\kappa}$ and shifts with $\gamma_c$ for varying $\langle z \rangle$. At high strain, ratios for all networks (irrespective of $\langle z \rangle$ and $\tilde{\kappa}$) converge to the affine result.}
\end{figure}

In our disordered network models, which deform nonaffinely and always possess some anisotropy, normalizing the measured value of $N_1$ by the Lodge-Meissner value ($\sigma_{xz}\gamma$) yields a quantitative measure of the degree to which the network behaves as an affinely deforming isotropic material. Since $\sigma_{xz}\gamma$ is always positive, this quantity also indicates when $N_1$ is negative. For the remainder of this work, we report the normalized quantity $N_1/(\sigma_{xz}\gamma)$. For unrotated ($\phi = 0$) phantom triangular lattice networks, the dominant force chains at $\gamma_c$ are tensile and oriented with $\theta > \pi/4$ for small $\gamma_c$, predicting that $N_1$ will be negative in the limit of low $\tilde{\kappa}$. With increasing $\gamma_c$, i.e. decreasing $z$, the force chains should develop with orientations closer to the principal strain axis, so decreasing $z$ should bring $N_1$ closer to the Lodge-Meissner value.  In Fig.\ \ref{fig7}a, we plot $N_1/(\sigma_{xz}\gamma)$ as a function of strain for phantom triangular lattice-based networks with $\phi = 0$, small $\tilde{\kappa}$, and varying $z$, demonstrating that these exhibit a negative peak in $N_1$ at the $z$-dependent critical strain, corresponding to the highly anisotropic force chains with $\theta > \pi/4$ shown in Fig.\ \ref{fig6}d. As predicted, the magnitude of this peak decreases as $z$ decreases and $\gamma_c$ increases, as the applied strain causes the principal stress axis to approach the principal strain axis. Nevertheless, even relatively high $\gamma_c$ values yield an anomalous downward peak in $\gamma_c$, indicating that these networks become maximally anisotropic at the critical strain. At large strains, of order $1$ or greater, $N_1\approx\sigma_{xz}\gamma$ for all networks, as the deformation becomes increasingly affine above the critical strain and the principal stress axis approaches the principal strain axis. 

We further demonstrate in Fig.\ \ref{fig7}b that the peak is related to the critical strain-stiffening transition by showing that, in unrotated phantom triangular networks with constant $z$ and varying $\tilde{\kappa}$, the peak becomes sharper in the $\tilde{\kappa}\to0$ limit.  With increasing $\tilde{\kappa}$, the deformation becomes increasingly affine, so the peak disappears and $N_1$ grows increasingly positive. With $\tilde{\kappa}\to\infty$, the disordered network's response approaches that of the corresponding affinely deforming \textit{undiluted} triangular lattice, which actually yields $N_1/(\sigma_{xz}\gamma)>1$ for $\phi = 0$ due to its inherent angular anisotropy. Given the rotational symmetry of the triangular lattice for rotations of $n\pi/3$, we expect that, beyond small differences due to random dilution, any angular anisotropy-related mechanical behavior of the lattice should be similar for initial rotations $\phi= n\pi/6$ where $n$ is even, whereas the opposite behavior should occur for odd $n$. For intermediate angles, we should observe a transition between these two cases. In Fig.\ \ref{fig6}a, we show the response for the full undiluted triangular lattice, as well as that of diluted phantom triangular networks with varying $\tilde{\kappa}$, with $\phi = 0$, in comparison to the corresponding curves for the ``opposite'' initial orientation $\phi = \pi/6$. We see that, for the full and diluted networks, $N_1/(\sigma_{xz}\gamma)$ essentially flips about the Lodge-Meissner value of 1 when the initial lattice is rotated by $\pi/6$. That the peak for the low-$\tilde{\kappa}$ case flips in sign is supportive of the idea that the orientation of the dominant force chains, shown in Fig.\ \ref{fig6}d-e, controls the sign and magnitude of $N_1$ relative to $\sigma_{xz}\gamma$. In Fig.\ \ref{fig6}b, we show $N_1/(\sigma_{xz}\gamma)$ for the full range of $\phi$ in the low-$\tilde{\kappa}$ case, demonstrating the smooth transition between the aforementioned extremes for rotations of $n\pi/6$. If an angular average is taken, the Lodge-Meissner relation is satisfied. It is interesting to note that, even for the phantom diluted triangular lattice, certain intermediate rotations should approximately satisfy the Lodge-Meissner relation at the critical strain as long as the dominant force chains, and thus the principal stress axis, are parallel to the principal extension axis. Phantom FCC networks, which also exhibit angular anisotropy, show qualitatively similar behavior, with a downward peak in $N_1/(\sigma_{xz}\gamma)$ for $\phi = 0$.

\begin{figure}[ht]
	\includegraphics[width=1.0\columnwidth]{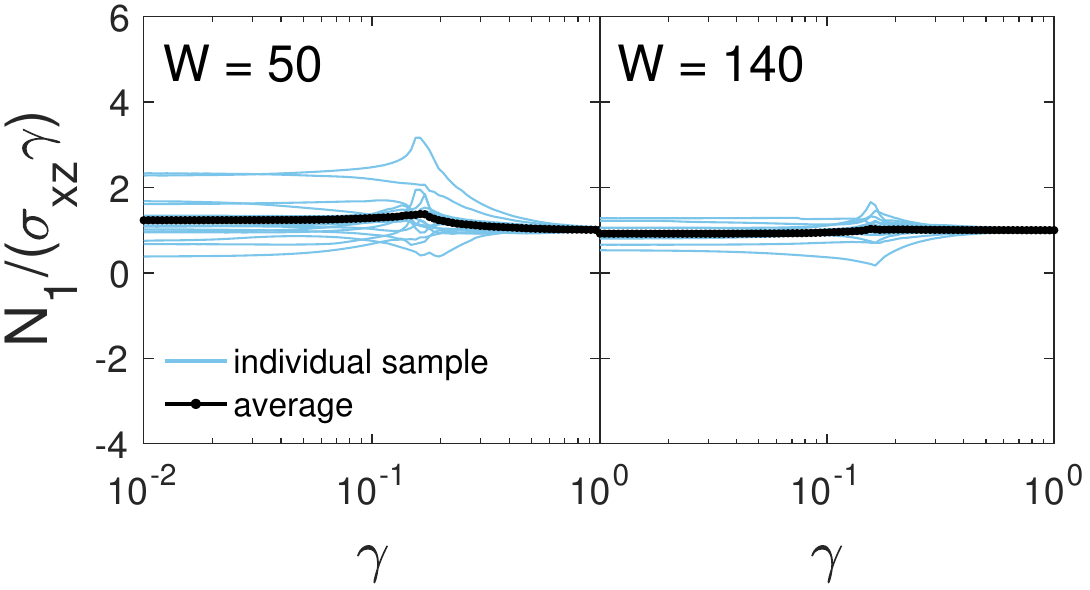}
	\caption{\label{fig8} Off-lattice networks show signatures of anisotropy in $N_1$ at the critical strain, but these deviations appear to average out in the limit of large system sizes or when averaged over many samples. For 2D packing-derived networks with $z = 3.3$ and  $\tilde{\kappa} = 10^{-6}$, we observe a decrease in the magnitude of the deviations of $N_1$ from the Lodge-Meissner relation with increasing system size.}
\end{figure}

For off-lattice networks with no long-range order, including Mikado and 2D/3D PD networks, the force chains that develop at the critical strain still occur with a directional bias towards the principal extension axis, but the lack of an underlying lattice structure means that they exhibit no orientational bias above or below the principal extension axis. Nevertheless, the highly heterogeneous and branched nature of these networks means that even for relatively large system sizes, some samples do exhibit deviation from the Lodge-Meissner relation at the critical strain. In Fig.\ \ref{fig8}a, we show that $N_1/(\sigma_{xz}\gamma)$ exhibits anomalous behavior with a peak at the critical strain for certain samples for small ($W = 50$, $N = 2500$ nodes) 2D packing-derived networks, indicating that this effect can occur in off-lattice networks. Averaging over an ensemble of initial network structures, the Lodge-Meissner relation is approximately satisfied. We show in Fig.\ \ref{fig8}b, that larger networks ($W = 140$, $N = 19600$ nodes) still exhibit anomalous peaks at the critical strain, but that these are typically lower in magnitude than those observed in smaller systems. The deviation from the LM relation in the bending-dominated regime appears to decrease with increasing system size as well. For a given network, the dominant force chains arise along the network's ``shortest paths`` \cite{Amuasi2015} consisting of connected bonds oriented close to the principal extension axis at a given strain, which have some excess length for $\gamma < \gamma_c$. The critical strain corresponds to the strain at which, in the $\tilde{\kappa} = 0$ limit, one or more of these shortest paths can no longer rearrange without the stretching of their constituent bonds. Thus, the structure of the force chain network and the resulting value of $N_1/(\sigma_{xz}\gamma)$, is determined at the critical strain by the orientations of these (initially randomly oriented) paths. 

While we do observe that individual samples typically closely approximate the Lodge-Meissner relation, it is unsurprising that finite-sized systems occasionally show anomalous behavior at the critical strain, as a consequence of the finite chance of some angular bias of the force chain network away from the principal extension axis. In the thermodynamic limit, the Lodge-Meissner relation should be satisfied even at the critical strain for individual networks, as increasing the system size should increase the likelihood that the system can ``find'' shortest paths close to the principal strain axis. In other words, deviation from the Lodge-Meissner relation requires a preferential orientation of the principal stress axis above (or below) the principal extension axis, which can only occur due to some underlying bond orientation bias in the initial network structure. For off-lattice models like packing-derived networks with no long-ranged structural anisotropy, such a preferential orientation is not possible in the limit of large system sizes, so the Lodge-Meissner relation is satisfied.  We observe the same behavior for 3D PD networks and Mikado networks as in 2D PD networks. It is worth noting that in off-lattice networks, like in the lattice-based networks, one can cause $N_1/(\sigma_{xz}\gamma)$ to flip about the Lodge-Meissner value by appropriately rotating the initial structure, and averaging over all possible initial orientations removes any deviation from Lodge-Meissner. 

\begin{figure}[ht]
	\includegraphics[width=1.0\columnwidth]{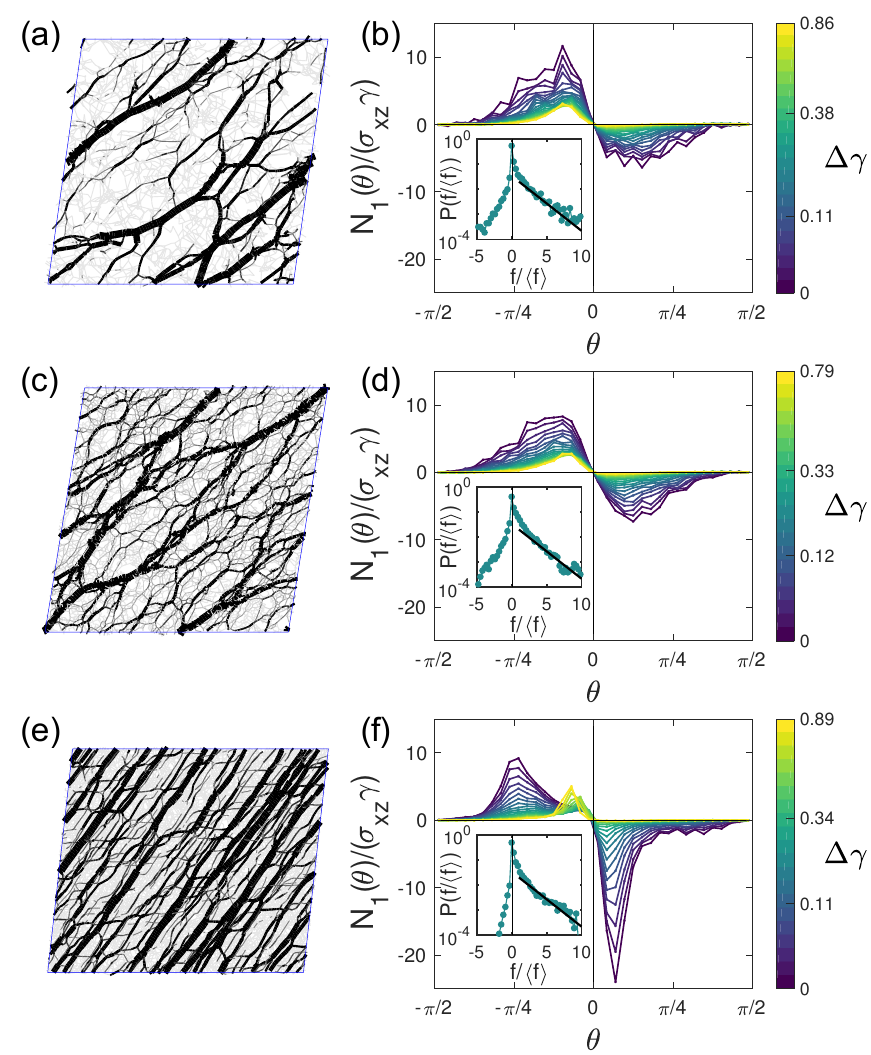}
	\caption{\label{fig9} At the critical strain in the limit of $\tilde{\kappa} = 0$, a system-spanning network of force chains develops that enables the network to bear finite stress. The angular orientation of this force network's constitutent bonds determines the sign of $N_1$. Here, we show representative force chains for central force ($\tilde{\kappa} = 0$) networks at the critical strain for (a) a Mikado network with $z = 3.3$, (b) a packing-derived network with $W = 100$ and $z = 3.4$, and (e) a phantomized triangular network with $z = 3.6$. We also show the corresponding distributions of $N_1(\theta)$, the contribution to $N_1$ from bonds oriented with a given angle $\theta$ relative to $\pi/4$, normalized by $\sigma_{xz}\gamma$ for varying $\Delta\gamma=\gamma-\gamma_c$. The integral of $N_1(\theta)$ over $\theta\in[-\pi/2,\pi/2]$ yields $N_1$. Hence, the relative areas of the positive and negative portions of the curve for a given $\gamma$ indicates the sign of $N_1$. The dominant contributions to $N_1$ are from bonds under tension. Insets: At the critical strain, the probability distribution of bond tension $f$, where $f_{ij} = \mu(l_{ij}/l_{ij,0}-1)$, normalized by the mean tensile force $\langle f\rangle=\mathrm{mean}(f(f>0))$, exhibits an exponential tail. The black solid lines corresponds to $P(f/\langle f \rangle)\propto\mathrm{exp}(-\beta(f/\langle f \rangle-1))$, with $\beta = 0.5$.}
\end{figure}

To emphasize the dependence of the value of $N_1$ on the highly heterogeneous force chain network structure at the critical strain, we show examples of force chains for Mikado, 2D PD, and unrotated phantom triangular networks with $\tilde{\kappa} = 0$ at the critical strain in Fig.\ \ref{fig9}. Mikado and PD networks show randomly branched force chains with a directional bias towards the principal extension axis, whereas the unrotated phantom triangular network shows the expected force chains oriented above the principal extension axis (and above $\theta_b = \pi/4$).  Additionally, we compute the distribution of contributions to $N_1$ due to bonds oriented with angle $\theta = \theta_b - \pi/4$, normalized by $\sigma_{xz}\gamma$, for each network structure as a function of $\Delta\gamma = \gamma - \gamma_c$, also shown in Fig.\ \ref{fig9}. Integrating these distributions over $\theta$ yields $N_1/(\sigma_{xz}\gamma)$ as a function of strain. At large strains, the networks all show very similar behavior, with primarily positive contributions to $N_1$ coming from primarily tensile bonds oriented close to the principal extension axis, below $\theta_b = \pi/4$, and with the total contribution satisfying the Lodge-Meissner relation. 

At the critical strain, however, the value of $N_1/(\sigma_{xz}\gamma)$ is determined by the balance of very large positive and negative contributions from bonds oriented above and below $\theta_b = \pi/4$. For the Mikado and PD networks shown, these positive and negative contributions are similar in magnitude at $\gamma_c$, but for the unrotated phantom triangular lattice, the negative contribution at $\gamma_c$ significantly outweighs the positive contribution, yielding the observed negative peak in $N_1/(\sigma_{xz}\gamma)$ vs. $\gamma$. The signficant heterogeneity of the force chain network is evident in noisy nature of these distributions at $\gamma_c$. We additionally plot, as insets in Fig.\ \ref{fig9}, the corresponding bond force distributions $P(f/\langle f \rangle)$ at $\gamma_c$, where $f > 0$ corresponds to tension and the average $\langle f \rangle$ is taken only over bonds under tension. Similar to observations of compressive force distributions in granular packings \cite{Radjai1996, OHern2001, Majmudar2005}, frictional forces in shear-thickening suspensions \cite{Melrose2004}, and tensile forces in polymer crazes \cite{Rottler2002}, we observe that the large ($f > \langle f \rangle$) tensile forces in our networks are, at the point of marginal stability i.e. $\gamma_c$, approximately exponentially distributed. To emphasize this, we show that the large forces can be approximated by the distribution $P(f/\langle f \rangle)\propto\mathrm{exp}(-\beta(f/\langle f \rangle-1))$. We find $\beta = 0.5$ appears to reasonably describe the distributions for the networks shown here. We also note that the compressive forces appear to exhibit an exponential tail as well, although they decay faster than the tensile forces. In a network of rope-like bonds or bucklable individual bonds with $\tilde{\kappa} = 0$, there would be no compressive forces. These distributions emphasize that tensile forces dominate at the critical strain. 

\section*{Summary and discussion}

In this work, we have shown the general scaling behavior of the normal stresses in the vicinity of the strain-driven stiffening transition for athermal semiflexible polymer networks, demonstrating that both the axial component $\sigma_{zz}$ and azimuthal component $\sigma_{xx}$ are quadratic in strain and proportional to the polymer bending rigidity $\tilde{\kappa}$ for $\gamma < \gamma_c$ but increase dramatically at the critical strain, such that both become proportional to the polymer stretching modulus $\mu$ for $\gamma > \gamma_c$. Additionally, we note that the critical strain coincides with the development of a heterogeneous network of primarily tensile force chains, similar to the compressive force chains observed in granular packings and frictional force chains observed in shear thickening suspensions. Along with the development of this force chain network, we observe a peak in the ratio of the axial normal stress to the shear stress (previously observed in prior work \cite{Conti2009, Kang2009}), which we show is a signature of the critical strain that becomes sharper with decreasing $\tilde{\kappa}$. For networks with low $\tilde{\kappa}$, we observe that this ratio appears to behave like the corresponding ratio for the affine rope network model, highlighting the primarily tensile nature of the stress-bearing force chain network. These observations possibly explain prior evidence that more flexible fibrin networks exhibit a sharper peak in $\sigma_{zz}/\sigma_{xz}$ than stiffer fibrin networks \cite{Kang2009}. 

Further, we observe that the highly anisotropic and heterogeneous structure of the force chain network that develops at the critical strain results in deviation of the first normal stress difference $N_1 = \sigma_{xx} - \sigma_{zz}$ from the Lodge-Meissner relation $N_1 = \sigma_{xz}\gamma$, particularly for networks with significant angular anisotropy (i.e. lattice-based models). This deviation from the Lodge-Meissner relation results from a difference in orientation of the principal stress axis from the principal extension axis, which in lattice-based models results from force chains at $\gamma_c$ developing primarily along whichever of the transformed initial lattice vector directions is closest to the principal strain axis. For lattice orientations in which the principal stress and strain axes do not align, we observe a peak in $N_1/(\sigma_{xz}\gamma)$ at the critical strain, consistent with the observation that these force chains are most anisotropic at the critical strain, and we show that appropriately rotating the lattice changes the sign of the peak. These results suggest that one can control the sign and magnitude of $N_1$ by modifying the network structure, similar to recent work showing that networks can be made auxetic by selectively pruning bonds \cite{Reid2018}. We observe that similar but typically smaller peaks in $N_1/(\sigma_{xz}\gamma)$ at $\gamma_c$ can also occur in off-lattice models, which lack long-range order and develop more random, branched force chain networks than lattice-based networks. While the force chain networks in off-lattice models are, on average, oriented along the principal strain axis, deviation from Lodge-Meissner is observed for finite systems at the critical strain and results from small imbalances between contributions to $N_1$ from bonds oriented on either side of the principal extension axis. Our results suggest that, in the thermodynamic limit, semiflexible networks with no long-range angular anisotropy (e.g. off-lattice models with $W \to \infty$) should satisfy the Lodge-Meissner relation, even at the critical strain.

This suggests that any observed deviation from the Lodge-Meissner relation in experimental measurements could serve as an indication of anisotropy in the network structure. For relatively isotropic biopolymer gels in which the sample size is much larger than the mesh size, we expect $N_1=\sigma_{xz}\gamma$, meaning that $N_1$, i.e. the measured normal stress on short timescales, can be expected to be positive. This is in agreement with experimental measurements of $N_1$ measured for fibrin gels at high frequencies \cite{DeCagny2016}. We  note that prior dynamic studies of spring networks have shown that viscous damping reduces nonaffinity at high frequencies \cite{Yucht2013, Dennison2016a}, which we expect to further reduce deviation from the Lodge-Meissner relation in this limit.

Finally, we report force probability distributions for networks at the critical strain in the limit of $\tilde{\kappa} = 0$, showing that the dominant forces at $\gamma_c$ are tensile, with additional evidence of an exponential tail in the large force probability distribution. Similar force probability distributions have been measured for other fragile or marginally stable systems, including compressive force networks in granular packings at the jamming point \cite{Radjai1996, OHern2001, Majmudar2005}, transient frictional force networks in sheared granular suspensions \cite{Melrose2004}, and force networks in polymer crazes \cite{Rottler2002}. Future work will be necessary to characterize these force networks and their implications in the strain-driven stiffening transition.

\begin{acknowledgments}
 This work was supported in part by the National Science Foundation Division of Materials Research (Grant DMR-1826623) and the National Science Foundation Center for Theoretical Biological Physics (Grant PHY-1427654).
\end{acknowledgments}


\begin{thebibliography}{58}%
\makeatletter
\providecommand \@ifxundefined [1]{%
 \@ifx{#1\undefined}
}%
\providecommand \@ifnum [1]{%
 \ifnum #1\expandafter \@firstoftwo
 \else \expandafter \@secondoftwo
 \fi
}%
\providecommand \@ifx [1]{%
 \ifx #1\expandafter \@firstoftwo
 \else \expandafter \@secondoftwo
 \fi
}%
\providecommand \natexlab [1]{#1}%
\providecommand \enquote  [1]{``#1''}%
\providecommand \bibnamefont  [1]{#1}%
\providecommand \bibfnamefont [1]{#1}%
\providecommand \citenamefont [1]{#1}%
\providecommand \href@noop [0]{\@secondoftwo}%
\providecommand \href [0]{\begingroup \@sanitize@url \@href}%
\providecommand \@href[1]{\@@startlink{#1}\@@href}%
\providecommand \@@href[1]{\endgroup#1\@@endlink}%
\providecommand \@sanitize@url [0]{\catcode `\\12\catcode `\$12\catcode
  `\&12\catcode `\#12\catcode `\^12\catcode `\_12\catcode `\%12\relax}%
\providecommand \@@startlink[1]{}%
\providecommand \@@endlink[0]{}%
\providecommand \url  [0]{\begingroup\@sanitize@url \@url }%
\providecommand \@url [1]{\endgroup\@href {#1}{\urlprefix }}%
\providecommand \urlprefix  [0]{URL }%
\providecommand \Eprint [0]{\href }%
\providecommand \doibase [0]{http://dx.doi.org/}%
\providecommand \selectlanguage [0]{\@gobble}%
\providecommand \bibinfo  [0]{\@secondoftwo}%
\providecommand \bibfield  [0]{\@secondoftwo}%
\providecommand \translation [1]{[#1]}%
\providecommand \BibitemOpen [0]{}%
\providecommand \bibitemStop [0]{}%
\providecommand \bibitemNoStop [0]{.\EOS\space}%
\providecommand \EOS [0]{\spacefactor3000\relax}%
\providecommand \BibitemShut  [1]{\csname bibitem#1\endcsname}%
\let\auto@bib@innerbib\@empty
\bibitem [{\citenamefont {Poynting}(1909)}]{Poynting1909}%
  \BibitemOpen
  \bibfield  {author} {\bibinfo {author} {\bibfnamefont {J.~H.}\ \bibnamefont
  {Poynting}},\ }\href
  {http://rspa.royalsocietypublishing.org/content/82/557/546.abstract}
  {\bibfield  {journal} {\bibinfo  {journal} {Proceedings of the Royal Society
  of London. Series A}\ }\textbf {\bibinfo {volume} {82}},\ \bibinfo {pages}
  {546} (\bibinfo {year} {1909})}\BibitemShut {NoStop}%
\bibitem [{\citenamefont {Poynting}(1913)}]{Poynting1913}%
  \BibitemOpen
  \bibfield  {author} {\bibinfo {author} {\bibfnamefont {J.~H.}\ \bibnamefont
  {Poynting}},\ }\href@noop {} {\bibfield  {journal} {\bibinfo  {journal} {The
  India-Rubber Journal}\ ,\ \bibinfo {pages} {6}} (\bibinfo {year}
  {1913})}\BibitemShut {NoStop}%
\bibitem [{\citenamefont {Larson}(1999)}]{Larson1999}%
  \BibitemOpen
  \bibfield  {author} {\bibinfo {author} {\bibfnamefont {R.~G.}\ \bibnamefont
  {Larson}},\ }\href@noop {} {\emph {\bibinfo {title} {{The Structure and
  Rheology of Complex Fluids}}}}\ (\bibinfo  {publisher} {Oxford University
  Press},\ \bibinfo {address} {New York},\ \bibinfo {year} {1999})\BibitemShut
  {NoStop}%
\bibitem [{\citenamefont {Lodge}\ and\ \citenamefont
  {Meissner}(1972)}]{Lodge1972}%
  \BibitemOpen
  \bibfield  {author} {\bibinfo {author} {\bibfnamefont {A.~S.}\ \bibnamefont
  {Lodge}}\ and\ \bibinfo {author} {\bibfnamefont {J.}~\bibnamefont
  {Meissner}},\ }\href {\doibase 10.1007/BF01974779} {\bibfield  {journal}
  {\bibinfo  {journal} {Rheologica Acta}\ }\textbf {\bibinfo {volume} {11}},\
  \bibinfo {pages} {351} (\bibinfo {year} {1972})}\BibitemShut {NoStop}%
\bibitem [{\citenamefont {Janmey}\ \emph {et~al.}(2007)\citenamefont {Janmey},
  \citenamefont {McCormick}, \citenamefont {Rammensee}, \citenamefont {Leight},
  \citenamefont {Georges},\ and\ \citenamefont {MacKintosh}}]{Janmey2007}%
  \BibitemOpen
  \bibfield  {author} {\bibinfo {author} {\bibfnamefont {P.~A.}\ \bibnamefont
  {Janmey}}, \bibinfo {author} {\bibfnamefont {M.~E.}\ \bibnamefont
  {McCormick}}, \bibinfo {author} {\bibfnamefont {S.}~\bibnamefont
  {Rammensee}}, \bibinfo {author} {\bibfnamefont {J.~L.}\ \bibnamefont
  {Leight}}, \bibinfo {author} {\bibfnamefont {P.~C.}\ \bibnamefont {Georges}},
  \ and\ \bibinfo {author} {\bibfnamefont {F.~C.}\ \bibnamefont {MacKintosh}},\
  }\href {\doibase 10.1038/nmat1810} {\bibfield  {journal} {\bibinfo  {journal}
  {Nature Materials}\ }\textbf {\bibinfo {volume} {6}},\ \bibinfo {pages} {48}
  (\bibinfo {year} {2007})}\BibitemShut {NoStop}%
\bibitem [{\citenamefont {Kang}\ \emph {et~al.}(2009)\citenamefont {Kang},
  \citenamefont {Wen}, \citenamefont {Janmey}, \citenamefont {Tang},
  \citenamefont {Conti},\ and\ \citenamefont {MacKintosh}}]{Kang2009}%
  \BibitemOpen
  \bibfield  {author} {\bibinfo {author} {\bibfnamefont {H.}~\bibnamefont
  {Kang}}, \bibinfo {author} {\bibfnamefont {Q.}~\bibnamefont {Wen}}, \bibinfo
  {author} {\bibfnamefont {P.~A.}\ \bibnamefont {Janmey}}, \bibinfo {author}
  {\bibfnamefont {J.~X.}\ \bibnamefont {Tang}}, \bibinfo {author}
  {\bibfnamefont {E.}~\bibnamefont {Conti}}, \ and\ \bibinfo {author}
  {\bibfnamefont {F.~C.}\ \bibnamefont {MacKintosh}},\ }\href {\doibase
  10.1021/jp807749f} {\bibfield  {journal} {\bibinfo  {journal} {Journal of
  Physical Chemistry B}\ }\textbf {\bibinfo {volume} {113}},\ \bibinfo {pages}
  {3799} (\bibinfo {year} {2009})}\BibitemShut {NoStop}%
\bibitem [{\citenamefont {Heussinger}\ \emph {et~al.}(2007)\citenamefont
  {Heussinger}, \citenamefont {Schaefer},\ and\ \citenamefont
  {Frey}}]{Heussinger2007}%
  \BibitemOpen
  \bibfield  {author} {\bibinfo {author} {\bibfnamefont {C.}~\bibnamefont
  {Heussinger}}, \bibinfo {author} {\bibfnamefont {B.}~\bibnamefont
  {Schaefer}}, \ and\ \bibinfo {author} {\bibfnamefont {E.}~\bibnamefont
  {Frey}},\ }\href {\doibase 10.1103/PhysRevE.76.031906} {\bibfield  {journal}
  {\bibinfo  {journal} {Physical Review E}\ }\textbf {\bibinfo {volume} {76}},\
  \bibinfo {pages} {1} (\bibinfo {year} {2007})}\BibitemShut {NoStop}%
\bibitem [{\citenamefont {Conti}\ and\ \citenamefont
  {MacKintosh}(2009)}]{Conti2009}%
  \BibitemOpen
  \bibfield  {author} {\bibinfo {author} {\bibfnamefont {E.}~\bibnamefont
  {Conti}}\ and\ \bibinfo {author} {\bibfnamefont {F.~C.}\ \bibnamefont
  {MacKintosh}},\ }\href {\doibase 10.1103/PhysRevLett.102.088102} {\bibfield
  {journal} {\bibinfo  {journal} {Physical Review Letters}\ }\textbf {\bibinfo
  {volume} {102}},\ \bibinfo {pages} {1} (\bibinfo {year} {2009})}\BibitemShut
  {NoStop}%
\bibitem [{\citenamefont {Cioroianu}\ and\ \citenamefont
  {Storm}(2013)}]{Cioroianu2013}%
  \BibitemOpen
  \bibfield  {author} {\bibinfo {author} {\bibfnamefont {A.~R.}\ \bibnamefont
  {Cioroianu}}\ and\ \bibinfo {author} {\bibfnamefont {C.}~\bibnamefont
  {Storm}},\ }\href {\doibase 10.1103/PhysRevE.88.052601} {\bibfield  {journal}
  {\bibinfo  {journal} {Physical Review E}\ }\textbf {\bibinfo {volume} {88}},\
  \bibinfo {pages} {1} (\bibinfo {year} {2013})}\BibitemShut {NoStop}%
\bibitem [{\citenamefont {Licup}\ \emph {et~al.}(2015)\citenamefont {Licup},
  \citenamefont {M{\"{u}}nster}, \citenamefont {Sharma}, \citenamefont
  {Sheinman}, \citenamefont {Jawerth}, \citenamefont {Fabry}, \citenamefont
  {Weitz},\ and\ \citenamefont {MacKintosh}}]{Licup2015}%
  \BibitemOpen
  \bibfield  {author} {\bibinfo {author} {\bibfnamefont {A.~J.}\ \bibnamefont
  {Licup}}, \bibinfo {author} {\bibfnamefont {S.}~\bibnamefont
  {M{\"{u}}nster}}, \bibinfo {author} {\bibfnamefont {A.}~\bibnamefont
  {Sharma}}, \bibinfo {author} {\bibfnamefont {M.}~\bibnamefont {Sheinman}},
  \bibinfo {author} {\bibfnamefont {L.~M.}\ \bibnamefont {Jawerth}}, \bibinfo
  {author} {\bibfnamefont {B.}~\bibnamefont {Fabry}}, \bibinfo {author}
  {\bibfnamefont {D.~A.}\ \bibnamefont {Weitz}}, \ and\ \bibinfo {author}
  {\bibfnamefont {F.~C.}\ \bibnamefont {MacKintosh}},\ }\href {\doibase
  10.1073/pnas.1504258112} {\bibfield  {journal} {\bibinfo  {journal}
  {Proceedings of the National Academy of Sciences}\ }\textbf {\bibinfo
  {volume} {112}},\ \bibinfo {pages} {201504258} (\bibinfo {year}
  {2015})}\BibitemShut {NoStop}%
\bibitem [{\citenamefont {Meng}\ and\ \citenamefont
  {Terentjev}(2016)}]{Meng2016}%
  \BibitemOpen
  \bibfield  {author} {\bibinfo {author} {\bibfnamefont {F.}~\bibnamefont
  {Meng}}\ and\ \bibinfo {author} {\bibfnamefont {E.~M.}\ \bibnamefont
  {Terentjev}},\ }\href {\doibase 10.1039/C6SM01029F} {\bibfield  {journal}
  {\bibinfo  {journal} {Soft Matter}\ }\textbf {\bibinfo {volume} {12}},\
  \bibinfo {pages} {6749} (\bibinfo {year} {2016})}\BibitemShut {NoStop}%
\bibitem [{\citenamefont {Licup}\ \emph {et~al.}(2016)\citenamefont {Licup},
  \citenamefont {Sharma},\ and\ \citenamefont {MacKintosh}}]{Licup2016}%
  \BibitemOpen
  \bibfield  {author} {\bibinfo {author} {\bibfnamefont {A.~J.}\ \bibnamefont
  {Licup}}, \bibinfo {author} {\bibfnamefont {A.}~\bibnamefont {Sharma}}, \
  and\ \bibinfo {author} {\bibfnamefont {F.~C.}\ \bibnamefont {MacKintosh}},\
  }\href {\doibase 10.1103/PhysRevE.93.012407} {\bibfield  {journal} {\bibinfo
  {journal} {Physical Review E}\ }\textbf {\bibinfo {volume} {93}},\ \bibinfo
  {pages} {1} (\bibinfo {year} {2016})}\BibitemShut {NoStop}%
\bibitem [{\citenamefont {Jansen}\ \emph {et~al.}(2018)\citenamefont {Jansen},
  \citenamefont {Licup}, \citenamefont {Sharma}, \citenamefont {Rens},
  \citenamefont {MacKintosh},\ and\ \citenamefont {Koenderink}}]{Jansen2018}%
  \BibitemOpen
  \bibfield  {author} {\bibinfo {author} {\bibfnamefont {K.~A.}\ \bibnamefont
  {Jansen}}, \bibinfo {author} {\bibfnamefont {A.~J.}\ \bibnamefont {Licup}},
  \bibinfo {author} {\bibfnamefont {A.}~\bibnamefont {Sharma}}, \bibinfo
  {author} {\bibfnamefont {R.}~\bibnamefont {Rens}}, \bibinfo {author}
  {\bibfnamefont {F.~C.}\ \bibnamefont {MacKintosh}}, \ and\ \bibinfo {author}
  {\bibfnamefont {G.~H.}\ \bibnamefont {Koenderink}},\ }\href {\doibase
  10.1016/j.bpj.2018.04.043} {\bibfield  {journal} {\bibinfo  {journal}
  {Biophysical Journal}\ }\textbf {\bibinfo {volume} {114}},\ \bibinfo {pages}
  {2665} (\bibinfo {year} {2018})}\BibitemShut {NoStop}%
\bibitem [{\citenamefont {{De Cagny}}\ \emph {et~al.}(2016)\citenamefont {{De
  Cagny}}, \citenamefont {Vos}, \citenamefont {Vahabi}, \citenamefont
  {Kurniawan}, \citenamefont {Doi}, \citenamefont {Koenderink}, \citenamefont
  {MacKintosh},\ and\ \citenamefont {Bonn}}]{DeCagny2016}%
  \BibitemOpen
  \bibfield  {author} {\bibinfo {author} {\bibfnamefont {H.~C.~G.}\
  \bibnamefont {{De Cagny}}}, \bibinfo {author} {\bibfnamefont {B.~E.}\
  \bibnamefont {Vos}}, \bibinfo {author} {\bibfnamefont {M.}~\bibnamefont
  {Vahabi}}, \bibinfo {author} {\bibfnamefont {N.~A.}\ \bibnamefont
  {Kurniawan}}, \bibinfo {author} {\bibfnamefont {M.}~\bibnamefont {Doi}},
  \bibinfo {author} {\bibfnamefont {G.~H.}\ \bibnamefont {Koenderink}},
  \bibinfo {author} {\bibfnamefont {F.~C.}\ \bibnamefont {MacKintosh}}, \ and\
  \bibinfo {author} {\bibfnamefont {D.}~\bibnamefont {Bonn}},\ }\href {\doibase
  10.1103/PhysRevLett.117.217802} {\bibfield  {journal} {\bibinfo  {journal}
  {Physical Review Letters}\ }\textbf {\bibinfo {volume} {117}},\ \bibinfo
  {pages} {1} (\bibinfo {year} {2016})}\BibitemShut {NoStop}%
\bibitem [{\citenamefont {Vahabi}\ \emph {et~al.}(2017)\citenamefont {Vahabi},
  \citenamefont {Vos}, \citenamefont {de~Cagny}, \citenamefont {Bonn},
  \citenamefont {Koenderink},\ and\ \citenamefont {MacKintosh}}]{Vahabi2017}%
  \BibitemOpen
  \bibfield  {author} {\bibinfo {author} {\bibfnamefont {M.}~\bibnamefont
  {Vahabi}}, \bibinfo {author} {\bibfnamefont {B.~E.}\ \bibnamefont {Vos}},
  \bibinfo {author} {\bibfnamefont {H.~C.~G.}\ \bibnamefont {de~Cagny}},
  \bibinfo {author} {\bibfnamefont {D.}~\bibnamefont {Bonn}}, \bibinfo {author}
  {\bibfnamefont {G.~H.}\ \bibnamefont {Koenderink}}, \ and\ \bibinfo {author}
  {\bibfnamefont {F.~C.}\ \bibnamefont {MacKintosh}},\ }\href@noop {}
  {\bibfield  {journal} {\bibinfo  {journal} {Physical Review E}\ }\textbf
  {\bibinfo {volume} {97}},\ \bibinfo {pages} {32418} (\bibinfo {year}
  {2017})}\BibitemShut {NoStop}%
\bibitem [{\citenamefont {Rivlin}(1948)}]{Rivlin1948}%
  \BibitemOpen
  \bibfield  {author} {\bibinfo {author} {\bibfnamefont {R.~S.}\ \bibnamefont
  {Rivlin}},\ }\href {\doibase 10.1098/rsta.1948.0024} {\bibfield  {journal}
  {\bibinfo  {journal} {Philosophical Transactions of the Royal Society A:
  Mathematical, Physical and Engineering Sciences}\ }\textbf {\bibinfo {volume}
  {241}},\ \bibinfo {pages} {379} (\bibinfo {year} {1948})}\BibitemShut
  {NoStop}%
\bibitem [{\citenamefont {Larson}\ and\ \citenamefont
  {Brenner}(1988)}]{Larson1988}%
  \BibitemOpen
  \bibfield  {author} {\bibinfo {author} {\bibfnamefont {R.~G.}\ \bibnamefont
  {Larson}}\ and\ \bibinfo {author} {\bibfnamefont {H.}~\bibnamefont
  {Brenner}},\ }\href {https://books.google.com/books?id=6uMgBQAAQBAJ} {\emph
  {\bibinfo {title} {{Constitutive Equations for Polymer Melts and Solutions:
  Butterworths Series in Chemical Engineering}}}},\ Butterworths Series in
  Chemical Engineering\ (\bibinfo  {publisher} {Elsevier Science},\ \bibinfo
  {year} {1988})\BibitemShut {NoStop}%
\bibitem [{\citenamefont {Sharma}\ \emph
  {et~al.}(2016{\natexlab{a}})\citenamefont {Sharma}, \citenamefont {Licup},
  \citenamefont {Jansen}, \citenamefont {Rens}, \citenamefont {Sheinman},
  \citenamefont {Koenderink},\ and\ \citenamefont {MacKintosh}}]{Sharma2016a}%
  \BibitemOpen
  \bibfield  {author} {\bibinfo {author} {\bibfnamefont {A.}~\bibnamefont
  {Sharma}}, \bibinfo {author} {\bibfnamefont {A.~J.}\ \bibnamefont {Licup}},
  \bibinfo {author} {\bibfnamefont {K.~A.}\ \bibnamefont {Jansen}}, \bibinfo
  {author} {\bibfnamefont {R.}~\bibnamefont {Rens}}, \bibinfo {author}
  {\bibfnamefont {M.}~\bibnamefont {Sheinman}}, \bibinfo {author}
  {\bibfnamefont {G.~H.}\ \bibnamefont {Koenderink}}, \ and\ \bibinfo {author}
  {\bibfnamefont {F.~C.}\ \bibnamefont {MacKintosh}},\ }\href {\doibase
  10.1038/nphys3628} {\bibfield  {journal} {\bibinfo  {journal} {Nature
  Physics}\ }\textbf {\bibinfo {volume} {12}},\ \bibinfo {pages} {584}
  (\bibinfo {year} {2016}{\natexlab{a}})}\BibitemShut {NoStop}%
\bibitem [{\citenamefont {Majmudar}\ and\ \citenamefont
  {Behringer}(2005)}]{Majmudar2005}%
  \BibitemOpen
  \bibfield  {author} {\bibinfo {author} {\bibfnamefont {T.~S.}\ \bibnamefont
  {Majmudar}}\ and\ \bibinfo {author} {\bibfnamefont {R.~P.}\ \bibnamefont
  {Behringer}},\ }\href {\doibase 10.1038/nature03805} {\bibfield  {journal}
  {\bibinfo  {journal} {Nature}\ }\textbf {\bibinfo {volume} {435}},\ \bibinfo
  {pages} {1079} (\bibinfo {year} {2005})}\BibitemShut {NoStop}%
\bibitem [{\citenamefont {Voivret}\ \emph {et~al.}(2009)\citenamefont
  {Voivret}, \citenamefont {Radja{\"{i}}}, \citenamefont {Delenne},\ and\
  \citenamefont {{El Youssoufi}}}]{Voivret2009}%
  \BibitemOpen
  \bibfield  {author} {\bibinfo {author} {\bibfnamefont {C.}~\bibnamefont
  {Voivret}}, \bibinfo {author} {\bibfnamefont {F.}~\bibnamefont
  {Radja{\"{i}}}}, \bibinfo {author} {\bibfnamefont {J.~Y.}\ \bibnamefont
  {Delenne}}, \ and\ \bibinfo {author} {\bibfnamefont {M.~S.}\ \bibnamefont
  {{El Youssoufi}}},\ }\href {\doibase 10.1103/PhysRevLett.102.178001}
  {\bibfield  {journal} {\bibinfo  {journal} {Physical Review Letters}\
  }\textbf {\bibinfo {volume} {102}},\ \bibinfo {pages} {2} (\bibinfo {year}
  {2009})}\BibitemShut {NoStop}%
\bibitem [{\citenamefont {Melrose}\ and\ \citenamefont
  {Ball}(2004)}]{Melrose2004}%
  \BibitemOpen
  \bibfield  {author} {\bibinfo {author} {\bibfnamefont {J.~R.}\ \bibnamefont
  {Melrose}}\ and\ \bibinfo {author} {\bibfnamefont {R.~C.}\ \bibnamefont
  {Ball}},\ }\href {\doibase 10.1122/1.1784784} {\bibfield  {journal} {\bibinfo
   {journal} {Journal of Rheology}\ }\textbf {\bibinfo {volume} {48}},\
  \bibinfo {pages} {961} (\bibinfo {year} {2004})}\BibitemShut {NoStop}%
\bibitem [{\citenamefont {Lin}\ \emph {et~al.}(2016)\citenamefont {Lin},
  \citenamefont {Ness}, \citenamefont {Cates}, \citenamefont {Sun},\ and\
  \citenamefont {Cohen}}]{Lin2016}%
  \BibitemOpen
  \bibfield  {author} {\bibinfo {author} {\bibfnamefont {N.~Y.~C.}\
  \bibnamefont {Lin}}, \bibinfo {author} {\bibfnamefont {C.}~\bibnamefont
  {Ness}}, \bibinfo {author} {\bibfnamefont {M.~E.}\ \bibnamefont {Cates}},
  \bibinfo {author} {\bibfnamefont {J.}~\bibnamefont {Sun}}, \ and\ \bibinfo
  {author} {\bibfnamefont {I.}~\bibnamefont {Cohen}},\ }\href
  {https://www.pnas.org/content/113/39/10774} {\bibfield  {journal} {\bibinfo
  {journal} {Proceedings of the National Academy of Sciences}\ }\textbf
  {\bibinfo {volume} {113}},\ \bibinfo {pages} {10774} (\bibinfo {year}
  {2016})}\BibitemShut {NoStop}%
\bibitem [{\citenamefont {Heussinger}\ and\ \citenamefont
  {Frey}(2007)}]{Heussinger2007a}%
  \BibitemOpen
  \bibfield  {author} {\bibinfo {author} {\bibfnamefont {C.}~\bibnamefont
  {Heussinger}}\ and\ \bibinfo {author} {\bibfnamefont {E.}~\bibnamefont
  {Frey}},\ }\href@noop {} {\bibfield  {journal} {\bibinfo  {journal} {European
  Physical Journal E}\ }\textbf {\bibinfo {volume} {24}},\ \bibinfo {pages}
  {47} (\bibinfo {year} {2007})}\BibitemShut {NoStop}%
\bibitem [{\citenamefont {Amuasi}\ \emph {et~al.}(2015)\citenamefont {Amuasi},
  \citenamefont {Heussinger}, \citenamefont {Vink},\ and\ \citenamefont
  {Zippelius}}]{Amuasi2015}%
  \BibitemOpen
  \bibfield  {author} {\bibinfo {author} {\bibfnamefont {H.~E.}\ \bibnamefont
  {Amuasi}}, \bibinfo {author} {\bibfnamefont {C.}~\bibnamefont {Heussinger}},
  \bibinfo {author} {\bibfnamefont {R.~L.}\ \bibnamefont {Vink}}, \ and\
  \bibinfo {author} {\bibfnamefont {A.}~\bibnamefont {Zippelius}},\ }\href@noop
  {} {\bibfield  {journal} {\bibinfo  {journal} {New Journal of Physics}\
  }\textbf {\bibinfo {volume} {17}} (\bibinfo {year} {2015})}\BibitemShut
  {NoStop}%
\bibitem [{\citenamefont {Liang}\ \emph {et~al.}(2016)\citenamefont {Liang},
  \citenamefont {Jones}, \citenamefont {Chen}, \citenamefont {Sun},\ and\
  \citenamefont {Jiao}}]{Liang2016}%
  \BibitemOpen
  \bibfield  {author} {\bibinfo {author} {\bibfnamefont {L.}~\bibnamefont
  {Liang}}, \bibinfo {author} {\bibfnamefont {C.}~\bibnamefont {Jones}},
  \bibinfo {author} {\bibfnamefont {S.}~\bibnamefont {Chen}}, \bibinfo {author}
  {\bibfnamefont {B.}~\bibnamefont {Sun}}, \ and\ \bibinfo {author}
  {\bibfnamefont {Y.}~\bibnamefont {Jiao}},\ }\href@noop {} {\bibfield
  {journal} {\bibinfo  {journal} {Physical Biology}\ }\textbf {\bibinfo
  {volume} {13}} (\bibinfo {year} {2016})}\BibitemShut {NoStop}%
\bibitem [{\citenamefont {Ronceray}\ \emph {et~al.}(2016)\citenamefont
  {Ronceray}, \citenamefont {Broedersz},\ and\ \citenamefont
  {Lenz}}]{Ronceray2016}%
  \BibitemOpen
  \bibfield  {author} {\bibinfo {author} {\bibfnamefont {P.}~\bibnamefont
  {Ronceray}}, \bibinfo {author} {\bibfnamefont {C.~P.}\ \bibnamefont
  {Broedersz}}, \ and\ \bibinfo {author} {\bibfnamefont {M.}~\bibnamefont
  {Lenz}},\ }\href {\doibase 10.1073/pnas.1514208113} {\bibfield  {journal}
  {\bibinfo  {journal} {Proceedings of the National Academy of Sciences}\
  }\textbf {\bibinfo {volume} {113}},\ \bibinfo {pages} {2827} (\bibinfo {year}
  {2016})}\BibitemShut {NoStop}%
\bibitem [{\citenamefont {Mari}\ \emph {et~al.}(2014)\citenamefont {Mari},
  \citenamefont {Seto}, \citenamefont {Morris},\ and\ \citenamefont
  {Denn}}]{Mari2014}%
  \BibitemOpen
  \bibfield  {author} {\bibinfo {author} {\bibfnamefont {R.}~\bibnamefont
  {Mari}}, \bibinfo {author} {\bibfnamefont {R.}~\bibnamefont {Seto}}, \bibinfo
  {author} {\bibfnamefont {J.~F.}\ \bibnamefont {Morris}}, \ and\ \bibinfo
  {author} {\bibfnamefont {M.~M.}\ \bibnamefont {Denn}},\ }\href {\doibase
  10.1122/1.4890747} {\bibfield  {journal} {\bibinfo  {journal} {Journal of
  Rheology}\ }\textbf {\bibinfo {volume} {58}},\ \bibinfo {pages} {1693}
  (\bibinfo {year} {2014})}\BibitemShut {NoStop}%
\bibitem [{\citenamefont {Cwalina}\ and\ \citenamefont
  {Wagner}(2014)}]{Cwalina2014}%
  \BibitemOpen
  \bibfield  {author} {\bibinfo {author} {\bibfnamefont {C.~D.}\ \bibnamefont
  {Cwalina}}\ and\ \bibinfo {author} {\bibfnamefont {N.~J.}\ \bibnamefont
  {Wagner}},\ }\href {\doibase 10.1122/1.4876935} {\bibfield  {journal}
  {\bibinfo  {journal} {Journal of Rheology}\ }\textbf {\bibinfo {volume}
  {58}},\ \bibinfo {pages} {949} (\bibinfo {year} {2014})}\BibitemShut
  {NoStop}%
\bibitem [{\citenamefont {Seto}\ and\ \citenamefont
  {Giusteri}(2018)}]{Seto2018}%
  \BibitemOpen
  \bibfield  {author} {\bibinfo {author} {\bibfnamefont {R.}~\bibnamefont
  {Seto}}\ and\ \bibinfo {author} {\bibfnamefont {G.~G.}\ \bibnamefont
  {Giusteri}},\ }\href {\doibase 10.1017/jfm.2018.743} {\bibfield  {journal}
  {\bibinfo  {journal} {Journal of Fluid Mechanics}\ }\textbf {\bibinfo
  {volume} {857}},\ \bibinfo {pages} {200} (\bibinfo {year}
  {2018})}\BibitemShut {NoStop}%
\bibitem [{\citenamefont {Thorpe}(1985)}]{Thorpe1985}%
  \BibitemOpen
  \bibfield  {author} {\bibinfo {author} {\bibfnamefont {M.~F.}\ \bibnamefont
  {Thorpe}},\ }\href {\doibase 10.1016/0022-3093(85)90056-0} {\bibfield
  {journal} {\bibinfo  {journal} {Journal of Non-Crystalline Solids}\ }\textbf
  {\bibinfo {volume} {76}},\ \bibinfo {pages} {109} (\bibinfo {year}
  {1985})}\BibitemShut {NoStop}%
\bibitem [{\citenamefont {Das}\ \emph {et~al.}(2007)\citenamefont {Das},
  \citenamefont {MacKintosh},\ and\ \citenamefont {Levine}}]{Das2007}%
  \BibitemOpen
  \bibfield  {author} {\bibinfo {author} {\bibfnamefont {M.}~\bibnamefont
  {Das}}, \bibinfo {author} {\bibfnamefont {F.~C.}\ \bibnamefont {MacKintosh}},
  \ and\ \bibinfo {author} {\bibfnamefont {A.~J.}\ \bibnamefont {Levine}},\
  }\href {\doibase 10.1103/PhysRevLett.99.038101} {\bibfield  {journal}
  {\bibinfo  {journal} {Physical Review Letters}\ }\textbf {\bibinfo {volume}
  {99}},\ \bibinfo {pages} {038101} (\bibinfo {year} {2007})}\BibitemShut
  {NoStop}%
\bibitem [{\citenamefont {Broedersz}\ and\ \citenamefont
  {MacKintosh}(2011)}]{Broedersz2011a}%
  \BibitemOpen
  \bibfield  {author} {\bibinfo {author} {\bibfnamefont {C.~P.}\ \bibnamefont
  {Broedersz}}\ and\ \bibinfo {author} {\bibfnamefont {F.~C.}\ \bibnamefont
  {MacKintosh}},\ }\href {\doibase 10.1039/c0sm01004a} {\bibfield  {journal}
  {\bibinfo  {journal} {Soft Matter}\ }\textbf {\bibinfo {volume} {7}},\
  \bibinfo {pages} {3186} (\bibinfo {year} {2011})}\BibitemShut {NoStop}%
\bibitem [{\citenamefont {Jacobs}\ and\ \citenamefont
  {Thorpe}(1996)}]{Jacobs1996}%
  \BibitemOpen
  \bibfield  {author} {\bibinfo {author} {\bibfnamefont {D.~J.}\ \bibnamefont
  {Jacobs}}\ and\ \bibinfo {author} {\bibfnamefont {M.~F.}\ \bibnamefont
  {Thorpe}},\ }\href@noop {} {\bibfield  {journal} {\bibinfo  {journal}
  {Physical Review E}\ }\textbf {\bibinfo {volume} {53}},\ \bibinfo {pages}
  {3682} (\bibinfo {year} {1996})}\BibitemShut {NoStop}%
\bibitem [{\citenamefont {Onck}\ \emph {et~al.}(2005)\citenamefont {Onck},
  \citenamefont {Koeman}, \citenamefont {van Dillen},\ and\ \citenamefont
  {van~der Giessen}}]{Onck2005}%
  \BibitemOpen
  \bibfield  {author} {\bibinfo {author} {\bibfnamefont {P.~R.}\ \bibnamefont
  {Onck}}, \bibinfo {author} {\bibfnamefont {T.}~\bibnamefont {Koeman}},
  \bibinfo {author} {\bibfnamefont {T.}~\bibnamefont {van Dillen}}, \ and\
  \bibinfo {author} {\bibfnamefont {E.}~\bibnamefont {van~der Giessen}},\
  }\href {\doibase 10.1103/PhysRevLett.95.178102} {\bibfield  {journal}
  {\bibinfo  {journal} {Physical Review Letters}\ }\textbf {\bibinfo {volume}
  {95}},\ \bibinfo {pages} {178102} (\bibinfo {year} {2005})}\BibitemShut
  {NoStop}%
\bibitem [{\citenamefont {Rens}\ \emph {et~al.}(2016)\citenamefont {Rens},
  \citenamefont {Vahabi}, \citenamefont {Licup}, \citenamefont {MacKintosh},\
  and\ \citenamefont {Sharma}}]{Rens2016}%
  \BibitemOpen
  \bibfield  {author} {\bibinfo {author} {\bibfnamefont {R.}~\bibnamefont
  {Rens}}, \bibinfo {author} {\bibfnamefont {M.}~\bibnamefont {Vahabi}},
  \bibinfo {author} {\bibfnamefont {A.~J.}\ \bibnamefont {Licup}}, \bibinfo
  {author} {\bibfnamefont {F.~C.}\ \bibnamefont {MacKintosh}}, \ and\ \bibinfo
  {author} {\bibfnamefont {A.}~\bibnamefont {Sharma}},\ }\href {\doibase
  10.1021/acs.jpcb.6b00259} {\bibfield  {journal} {\bibinfo  {journal} {Journal
  of Physical Chemistry B}\ }\textbf {\bibinfo {volume} {120}},\ \bibinfo
  {pages} {5831} (\bibinfo {year} {2016})}\BibitemShut {NoStop}%
\bibitem [{\citenamefont {Broedersz}\ \emph {et~al.}(2011)\citenamefont
  {Broedersz}, \citenamefont {Mao}, \citenamefont {Lubensky},\ and\
  \citenamefont {MacKintosh}}]{Broedersz2011}%
  \BibitemOpen
  \bibfield  {author} {\bibinfo {author} {\bibfnamefont {C.~P.}\ \bibnamefont
  {Broedersz}}, \bibinfo {author} {\bibfnamefont {X.}~\bibnamefont {Mao}},
  \bibinfo {author} {\bibfnamefont {T.~C.}\ \bibnamefont {Lubensky}}, \ and\
  \bibinfo {author} {\bibfnamefont {F.~C.}\ \bibnamefont {MacKintosh}},\ }\href
  {http://dx.doi.org/10.1038/nphys2127
  http://www.nature.com/nphys/journal/v7/n12/abs/nphys2127.html{\#}supplementary-information}
  {\bibfield  {journal} {\bibinfo  {journal} {Nature Physics}\ }\textbf
  {\bibinfo {volume} {7}},\ \bibinfo {pages} {983} (\bibinfo {year}
  {2011})}\BibitemShut {NoStop}%
\bibitem [{\citenamefont {Feng}\ and\ \citenamefont {Sen}(1984)}]{Feng1984}%
  \BibitemOpen
  \bibfield  {author} {\bibinfo {author} {\bibfnamefont {S.}~\bibnamefont
  {Feng}}\ and\ \bibinfo {author} {\bibfnamefont {P.~N.}\ \bibnamefont {Sen}},\
  }\href {\doibase 10.1103/PhysRevLett.52.216} {\bibfield  {journal} {\bibinfo
  {journal} {Physical Review Letters}\ }\textbf {\bibinfo {volume} {52}},\
  \bibinfo {pages} {216} (\bibinfo {year} {1984})}\BibitemShut {NoStop}%
\bibitem [{\citenamefont {Broedersz}\ \emph {et~al.}(2012)\citenamefont
  {Broedersz}, \citenamefont {Sheinman},\ and\ \citenamefont
  {MacKintosh}}]{Broedersz2012}%
  \BibitemOpen
  \bibfield  {author} {\bibinfo {author} {\bibfnamefont {C.~P.}\ \bibnamefont
  {Broedersz}}, \bibinfo {author} {\bibfnamefont {M.}~\bibnamefont {Sheinman}},
  \ and\ \bibinfo {author} {\bibfnamefont {F.~C.}\ \bibnamefont {MacKintosh}},\
  }\href {\doibase 10.1103/PhysRevLett.108.078102} {\bibfield  {journal}
  {\bibinfo  {journal} {Physical Review Letters}\ }\textbf {\bibinfo {volume}
  {108}},\ \bibinfo {pages} {078102} (\bibinfo {year} {2012})}\BibitemShut
  {NoStop}%
\bibitem [{\citenamefont {Wilhelm}\ and\ \citenamefont
  {Frey}(2003)}]{Wilhelm2003}%
  \BibitemOpen
  \bibfield  {author} {\bibinfo {author} {\bibfnamefont {J.}~\bibnamefont
  {Wilhelm}}\ and\ \bibinfo {author} {\bibfnamefont {E.}~\bibnamefont {Frey}},\
  }\href {\doibase 10.1103/PhysRevLett.91.108103} {\bibfield  {journal}
  {\bibinfo  {journal} {Physical Review Letters}\ }\textbf {\bibinfo {volume}
  {91}},\ \bibinfo {pages} {108103} (\bibinfo {year} {2003})}\BibitemShut
  {NoStop}%
\bibitem [{\citenamefont {Head}\ \emph {et~al.}(2003)\citenamefont {Head},
  \citenamefont {Levine},\ and\ \citenamefont {MacKintosh}}]{Head2003}%
  \BibitemOpen
  \bibfield  {author} {\bibinfo {author} {\bibfnamefont {D.~A.}\ \bibnamefont
  {Head}}, \bibinfo {author} {\bibfnamefont {A.~J.}\ \bibnamefont {Levine}}, \
  and\ \bibinfo {author} {\bibfnamefont {F.~C.}\ \bibnamefont {MacKintosh}},\
  }\href {\doibase 10.1103/PhysRevLett.91.108102} {\bibfield  {journal}
  {\bibinfo  {journal} {Physical Review Letters}\ }\textbf {\bibinfo {volume}
  {91}},\ \bibinfo {pages} {108102} (\bibinfo {year} {2003})}\BibitemShut
  {NoStop}%
\bibitem [{\citenamefont {Koeze}\ \emph {et~al.}(2016)\citenamefont {Koeze},
  \citenamefont {V{\aa}gberg}, \citenamefont {Tjoa},\ and\ \citenamefont
  {Tighe}}]{Koeze2016}%
  \BibitemOpen
  \bibfield  {author} {\bibinfo {author} {\bibfnamefont {D.~J.}\ \bibnamefont
  {Koeze}}, \bibinfo {author} {\bibfnamefont {D.}~\bibnamefont {V{\aa}gberg}},
  \bibinfo {author} {\bibfnamefont {B.~B.}\ \bibnamefont {Tjoa}}, \ and\
  \bibinfo {author} {\bibfnamefont {B.~P.}\ \bibnamefont {Tighe}},\ }\href@noop
  {} {\bibfield  {journal} {\bibinfo  {journal} {Europhysics Letters}\ }\textbf
  {\bibinfo {volume} {113}} (\bibinfo {year} {2016})}\BibitemShut {NoStop}%
\bibitem [{\citenamefont {{Van Hecke}}(2010)}]{VanHecke2010}%
  \BibitemOpen
  \bibfield  {author} {\bibinfo {author} {\bibfnamefont {M.}~\bibnamefont {{Van
  Hecke}}},\ }\href@noop {} {\bibfield  {journal} {\bibinfo  {journal} {Journal
  of Physics Condensed Matter}\ }\textbf {\bibinfo {volume} {22}} (\bibinfo
  {year} {2010})}\BibitemShut {NoStop}%
\bibitem [{\citenamefont {O'hern}\ \emph {et~al.}(2003)\citenamefont {O'hern},
  \citenamefont {Silbert},\ and\ \citenamefont {Nagel}}]{Ohern2003}%
  \BibitemOpen
  \bibfield  {author} {\bibinfo {author} {\bibfnamefont {C.}~\bibnamefont
  {O'hern}}, \bibinfo {author} {\bibfnamefont {L.}~\bibnamefont {Silbert}}, \
  and\ \bibinfo {author} {\bibfnamefont {S.}~\bibnamefont {Nagel}},\ }\href
  {http://dx.doi.org/10.1103/PhysRevE.68.011306{\%}5Cnfile:///Users/Lilian/Desktop/PhD
  Stuff/Papers/2003/O?hern/Phys. Rev. E 2003
  O?hern.pdf{\%}5Cnpapers://4bbe7dda-7fed-4d23-8c1c-c2c5fa8cce83/Paper/p4862}
  {\bibfield  {journal} {\bibinfo  {journal} {Physical Review E}\ }\textbf
  {\bibinfo {volume} {68}},\ \bibinfo {pages} {11306} (\bibinfo {year}
  {2003})}\BibitemShut {NoStop}%
\bibitem [{\citenamefont {Dagois-Bohy}\ \emph {et~al.}(2012)\citenamefont
  {Dagois-Bohy}, \citenamefont {Tighe}, \citenamefont {Simon}, \citenamefont
  {Henkes},\ and\ \citenamefont {{Van Hecke}}}]{Dagois-Bohy2012}%
  \BibitemOpen
  \bibfield  {author} {\bibinfo {author} {\bibfnamefont {S.}~\bibnamefont
  {Dagois-Bohy}}, \bibinfo {author} {\bibfnamefont {B.~P.}\ \bibnamefont
  {Tighe}}, \bibinfo {author} {\bibfnamefont {J.}~\bibnamefont {Simon}},
  \bibinfo {author} {\bibfnamefont {S.}~\bibnamefont {Henkes}}, \ and\ \bibinfo
  {author} {\bibfnamefont {M.}~\bibnamefont {{Van Hecke}}},\ }\href@noop {}
  {\bibfield  {journal} {\bibinfo  {journal} {Physical Review Letters}\
  }\textbf {\bibinfo {volume} {109}},\ \bibinfo {pages} {1} (\bibinfo {year}
  {2012})}\BibitemShut {NoStop}%
\bibitem [{\citenamefont {Arbabi}\ and\ \citenamefont
  {Sahimi}(1988)}]{Arbabi1988}%
  \BibitemOpen
  \bibfield  {author} {\bibinfo {author} {\bibfnamefont {S.}~\bibnamefont
  {Arbabi}}\ and\ \bibinfo {author} {\bibfnamefont {M.}~\bibnamefont
  {Sahimi}},\ }\href {\doibase 10.1103/PhysRevB.38.7173} {\bibfield  {journal}
  {\bibinfo  {journal} {Physical Review B}\ }\textbf {\bibinfo {volume} {38}},\
  \bibinfo {pages} {7173} (\bibinfo {year} {1988})}\BibitemShut {NoStop}%
\bibitem [{\citenamefont {Das}\ \emph {et~al.}(2012)\citenamefont {Das},
  \citenamefont {Quint},\ and\ \citenamefont {Schwarz}}]{Das2012}%
  \BibitemOpen
  \bibfield  {author} {\bibinfo {author} {\bibfnamefont {M.}~\bibnamefont
  {Das}}, \bibinfo {author} {\bibfnamefont {D.~A.}\ \bibnamefont {Quint}}, \
  and\ \bibinfo {author} {\bibfnamefont {J.~M.}\ \bibnamefont {Schwarz}},\
  }\href {https://doi.org/10.1371/journal.pone.0035939} {\bibfield  {journal}
  {\bibinfo  {journal} {PLoS ONE}\ }\textbf {\bibinfo {volume} {7}},\ \bibinfo
  {pages} {e35939} (\bibinfo {year} {2012})}\BibitemShut {NoStop}%
\bibitem [{\citenamefont {Sharma}\ \emph
  {et~al.}(2016{\natexlab{b}})\citenamefont {Sharma}, \citenamefont {Licup},
  \citenamefont {Rens}, \citenamefont {Vahabi}, \citenamefont {Jansen},
  \citenamefont {Koenderink},\ and\ \citenamefont {MacKintosh}}]{Sharma2016}%
  \BibitemOpen
  \bibfield  {author} {\bibinfo {author} {\bibfnamefont {A.}~\bibnamefont
  {Sharma}}, \bibinfo {author} {\bibfnamefont {A.~J.}\ \bibnamefont {Licup}},
  \bibinfo {author} {\bibfnamefont {R.}~\bibnamefont {Rens}}, \bibinfo {author}
  {\bibfnamefont {M.}~\bibnamefont {Vahabi}}, \bibinfo {author} {\bibfnamefont
  {K.~A.}\ \bibnamefont {Jansen}}, \bibinfo {author} {\bibfnamefont {G.~H.}\
  \bibnamefont {Koenderink}}, \ and\ \bibinfo {author} {\bibfnamefont {F.~C.}\
  \bibnamefont {MacKintosh}},\ }\href {\doibase 10.1103/PhysRevE.94.042407}
  {\bibfield  {journal} {\bibinfo  {journal} {Physical Review E}\ }\textbf
  {\bibinfo {volume} {94}},\ \bibinfo {pages} {042407} (\bibinfo {year}
  {2016}{\natexlab{b}})}\BibitemShut {NoStop}%
\bibitem [{\citenamefont {Lindstr{\"{o}}m}\ \emph {et~al.}(2010)\citenamefont
  {Lindstr{\"{o}}m}, \citenamefont {Vader}, \citenamefont {Kulachenko},\ and\
  \citenamefont {Weitz}}]{Lindstrom2010}%
  \BibitemOpen
  \bibfield  {author} {\bibinfo {author} {\bibfnamefont {S.~B.}\ \bibnamefont
  {Lindstr{\"{o}}m}}, \bibinfo {author} {\bibfnamefont {D.~A.}\ \bibnamefont
  {Vader}}, \bibinfo {author} {\bibfnamefont {A.}~\bibnamefont {Kulachenko}}, \
  and\ \bibinfo {author} {\bibfnamefont {D.~A.}\ \bibnamefont {Weitz}},\
  }\href@noop {} {\bibfield  {journal} {\bibinfo  {journal} {Physical Review
  E}\ }\textbf {\bibinfo {volume} {82}},\ \bibinfo {pages} {051905} (\bibinfo
  {year} {2010})}\BibitemShut {NoStop}%
\bibitem [{\citenamefont {Lees}\ and\ \citenamefont
  {Edwards}(1972)}]{Lees1972}%
  \BibitemOpen
  \bibfield  {author} {\bibinfo {author} {\bibfnamefont {A.~W.}\ \bibnamefont
  {Lees}}\ and\ \bibinfo {author} {\bibfnamefont {S.~F.}\ \bibnamefont
  {Edwards}},\ }\href {\doibase 10.1088/0022-3719/5/15/006} {\bibfield
  {journal} {\bibinfo  {journal} {Journal of Physics C: Solid State Physics}\
  }\textbf {\bibinfo {volume} {5}},\ \bibinfo {pages} {1921} (\bibinfo {year}
  {1972})}\BibitemShut {NoStop}%
\bibitem [{\citenamefont {Bitzek}\ \emph {et~al.}(2006)\citenamefont {Bitzek},
  \citenamefont {Koskinen}, \citenamefont {G{\"{a}}hler}, \citenamefont
  {Moseler},\ and\ \citenamefont {Gumbsch}}]{Bitzek}%
  \BibitemOpen
  \bibfield  {author} {\bibinfo {author} {\bibfnamefont {E.}~\bibnamefont
  {Bitzek}}, \bibinfo {author} {\bibfnamefont {P.}~\bibnamefont {Koskinen}},
  \bibinfo {author} {\bibfnamefont {F.}~\bibnamefont {G{\"{a}}hler}}, \bibinfo
  {author} {\bibfnamefont {M.}~\bibnamefont {Moseler}}, \ and\ \bibinfo
  {author} {\bibfnamefont {P.}~\bibnamefont {Gumbsch}},\ }\href {\doibase
  10.1103/PhysRevLett.97.170201} {\bibfield  {journal} {\bibinfo  {journal}
  {Physical Review Letters}\ }\textbf {\bibinfo {volume} {97}},\ \bibinfo
  {pages} {170201} (\bibinfo {year} {2006})}\BibitemShut {NoStop}%
\bibitem [{\citenamefont {Doi}\ and\ \citenamefont
  {Edwards}(1988)}]{DoiEdwards}%
  \BibitemOpen
  \bibfield  {author} {\bibinfo {author} {\bibfnamefont {M.}~\bibnamefont
  {Doi}}\ and\ \bibinfo {author} {\bibfnamefont {S.~F.}\ \bibnamefont
  {Edwards}},\ }\href@noop {} {\emph {\bibinfo {title} {{The Theory of Polymer
  Dynamics}}}},\ Vol.~\bibinfo {volume} {73}\ (\bibinfo  {publisher} {Oxford
  University Press},\ \bibinfo {year} {1988})\BibitemShut {NoStop}%
\bibitem [{\citenamefont {Maxwell}(1864)}]{Maxwell1864}%
  \BibitemOpen
  \bibfield  {author} {\bibinfo {author} {\bibfnamefont {J.~C.}\ \bibnamefont
  {Maxwell}},\ }\href@noop {} {\bibfield  {journal} {\bibinfo  {journal} {The
  London, Edinburgh, and Dublin Philosophical Magazine and Journal of Science}\
  }\textbf {\bibinfo {volume} {27}},\ \bibinfo {pages} {294} (\bibinfo {year}
  {1864})}\BibitemShut {NoStop}%
\bibitem [{\citenamefont {Radjai}\ \emph {et~al.}(1996)\citenamefont {Radjai},
  \citenamefont {Jean}, \citenamefont {Moreau},\ and\ \citenamefont
  {Roux}}]{Radjai1996}%
  \BibitemOpen
  \bibfield  {author} {\bibinfo {author} {\bibfnamefont {F.}~\bibnamefont
  {Radjai}}, \bibinfo {author} {\bibfnamefont {M.}~\bibnamefont {Jean}},
  \bibinfo {author} {\bibfnamefont {J.~J.}\ \bibnamefont {Moreau}}, \ and\
  \bibinfo {author} {\bibfnamefont {S.}~\bibnamefont {Roux}},\ }\href {\doibase
  10.1103/PhysRevLett.77.274} {\bibfield  {journal} {\bibinfo  {journal}
  {Physical Review Letters}\ }\textbf {\bibinfo {volume} {77}},\ \bibinfo
  {pages} {274} (\bibinfo {year} {1996})}\BibitemShut {NoStop}%
\bibitem [{\citenamefont {O'Hern}\ \emph {et~al.}(2001)\citenamefont {O'Hern},
  \citenamefont {Langer}, \citenamefont {Liu},\ and\ \citenamefont
  {Nagel}}]{OHern2001}%
  \BibitemOpen
  \bibfield  {author} {\bibinfo {author} {\bibfnamefont {C.~S.}\ \bibnamefont
  {O'Hern}}, \bibinfo {author} {\bibfnamefont {S.~A.}\ \bibnamefont {Langer}},
  \bibinfo {author} {\bibfnamefont {A.~J.}\ \bibnamefont {Liu}}, \ and\
  \bibinfo {author} {\bibfnamefont {S.~R.}\ \bibnamefont {Nagel}},\ }\href@noop
  {} {\bibfield  {journal} {\bibinfo  {journal} {Physical Review Letters}\
  }\textbf {\bibinfo {volume} {86}},\ \bibinfo {pages} {111} (\bibinfo {year}
  {2001})}\BibitemShut {NoStop}%
\bibitem [{\citenamefont {Rottler}\ and\ \citenamefont
  {Robbins}(2002)}]{Rottler2002}%
  \BibitemOpen
  \bibfield  {author} {\bibinfo {author} {\bibfnamefont {J.}~\bibnamefont
  {Rottler}}\ and\ \bibinfo {author} {\bibfnamefont {M.~O.}\ \bibnamefont
  {Robbins}},\ }\href@noop {} {\bibfield  {journal} {\bibinfo  {journal}
  {Physical Review Letters}\ }\textbf {\bibinfo {volume} {89}},\ \bibinfo
  {pages} {5} (\bibinfo {year} {2002})}\BibitemShut {NoStop}%
\bibitem [{\citenamefont {Reid}\ \emph {et~al.}(2018)\citenamefont {Reid},
  \citenamefont {Pashine}, \citenamefont {Wozniak}, \citenamefont {Jaeger},
  \citenamefont {Liu}, \citenamefont {Nagel},\ and\ \citenamefont
  {de~Pablo}}]{Reid2018}%
  \BibitemOpen
  \bibfield  {author} {\bibinfo {author} {\bibfnamefont {D.~R.}\ \bibnamefont
  {Reid}}, \bibinfo {author} {\bibfnamefont {N.}~\bibnamefont {Pashine}},
  \bibinfo {author} {\bibfnamefont {J.~M.}\ \bibnamefont {Wozniak}}, \bibinfo
  {author} {\bibfnamefont {H.~M.}\ \bibnamefont {Jaeger}}, \bibinfo {author}
  {\bibfnamefont {A.~J.}\ \bibnamefont {Liu}}, \bibinfo {author} {\bibfnamefont
  {S.~R.}\ \bibnamefont {Nagel}}, \ and\ \bibinfo {author} {\bibfnamefont
  {J.~J.}\ \bibnamefont {de~Pablo}},\ }\href {\doibase 10.1073/pnas.1717442115}
  {\bibfield  {journal} {\bibinfo  {journal} {Proceedings of the National
  Academy of Sciences}\ }\textbf {\bibinfo {volume} {115}},\ \bibinfo {pages}
  {E1384} (\bibinfo {year} {2018})}\BibitemShut {NoStop}%
\bibitem [{\citenamefont {Yucht}\ \emph {et~al.}(2013)\citenamefont {Yucht},
  \citenamefont {Sheinman},\ and\ \citenamefont {Broedersz}}]{Yucht2013}%
  \BibitemOpen
  \bibfield  {author} {\bibinfo {author} {\bibfnamefont {M.~G.}\ \bibnamefont
  {Yucht}}, \bibinfo {author} {\bibfnamefont {M.}~\bibnamefont {Sheinman}}, \
  and\ \bibinfo {author} {\bibfnamefont {C.~P.}\ \bibnamefont {Broedersz}},\
  }\href
  {http://dx.doi.org/10.1039/C3SM50177A{\%}5Cnhttp://pubs.rsc.org/en/content/articlepdf/2013/sm/c3sm50177a}
  {\bibfield  {journal} {\bibinfo  {journal} {Soft Matter}\ }\textbf {\bibinfo
  {volume} {9}},\ \bibinfo {pages} {7000} (\bibinfo {year} {2013})}\BibitemShut
  {NoStop}%
\bibitem [{\citenamefont {Dennison}\ and\ \citenamefont
  {Stark}(2016)}]{Dennison2016a}%
  \BibitemOpen
  \bibfield  {author} {\bibinfo {author} {\bibfnamefont {M.}~\bibnamefont
  {Dennison}}\ and\ \bibinfo {author} {\bibfnamefont {H.}~\bibnamefont
  {Stark}},\ }\href {\doibase 10.1103/PhysRevE.93.022605} {\bibfield  {journal}
  {\bibinfo  {journal} {Physical Review E}\ }\textbf {\bibinfo {volume} {93}},\
  \bibinfo {pages} {1} (\bibinfo {year} {2016})}\BibitemShut {NoStop}%
\end{thebibliography}

%

\end{document}